\documentclass[aps,prb,reprint,groupedaddress,showpacs]{revtex4-1}
\usepackage{graphicx,graphics}
\usepackage[table]{xcolor}
\usepackage{epstopdf}
\usepackage{amsmath,amssymb,amsfonts}
\usepackage{latexsym,verbatim}
\usepackage{bm}
\usepackage{bbold}
\usepackage{color}
\usepackage{ulem}
\usepackage[breaklinks=false,colorlinks,citecolor=blue,linkcolor=blue,urlcolor=blue]{hyperref}
\usepackage[abs]{overpic}
\usepackage{textcomp} 
\usepackage{mathtools}
\usepackage{braket}
\usepackage{comment}
\usepackage{tcolorbox}
\usepackage{soul}    

\begin{document}

\title{Vibrational response functions for multidimensional electronic spectroscopy \\ 
in non-adiabatic models}

\author{Filippo Troiani}
\affiliation{Centro S3, CNR-Istituto di Nanoscienze, I-41125 Modena, Italy}
\email{filippo.troiani@nano.cnr.it}

\begin{abstract}
The interplay of nuclear and electronic dynamics characterizes the multi-dimensional electronic spectra of various molecular and solid-state systems. Theoretically, the observable effect of such interplay can be accounted for by response functions. Here, we report analytical expressions for the response functions corresponding to a class of model systems. These are characterized by the coupling between the diabatic electronic states and the vibrational degrees of freedom resulting in linear displacements of the corresponding harmonic oscillators, and by non-adiabatic couplings between pairs of diabatic states. In order to derive the linear response functions, we first perform the Dyson expansion of the relevant propagators with respect to the non-adiabatic component of the Hamiltonian, then derive and expand with respect to the displacements the propagators at given interaction times, and finally provide analytical expressions for the time integrals that lead to the different contributions to the linear response function. The approach is then applied to the derivation of third-order response functions describing different physical processes: ground state bleaching, stimulated emission, excited state absorption and  double quantum coherence. Comparisons between the results obtained up to sixth order in the Dyson expansion and independent numerical calculation of the response functions provide an evidence of the series convergence in a few representative cases. 
\end{abstract}

\date{\today}

\maketitle

\section{Introduction}

Multidimensional coherent spectroscopy represents a powerful tool for investigating ultrafast dynamical processes occurring in molecular and solid-state systems\cite{Mukamel95a,Hamm11a,Scholes2017,Smallwood18a,Rozzi18a,Collini2021}. 
In fact, the dependence of the nonlinear spectra on multiple frequencies allows one to separate different and otherwise overlapping contributions, and to establish correlations between the observed excitation energies.

These processes often involve an interplay between electronic and vibrational degrees of freedom, which plays an important role in processes such as charge or energy transfer and determines the observed coherent beatings \cite{Chin2013,Falke14a,Romero2014,OReilly2014,DeSio16a,Thouin2019,Rafiq2021}. 
In a semiclassical representation of the system dynamics, ultrashort laser pulses induce impulsive transitions to different electronic states. This triggers the wave packet motion on the corresponding potential energy surfaces, with features that depend on the specific form of the electron-phonon coupling. In many cases of interest, such coupling is represented in terms of the linearly displaced-oscillator model, where each vibrational mode is represented as an independent harmonic oscillator, which undergoes an electronic-state dependent displacement of the origin \cite{Kumar01a,Egorova2007a,Mancal10a,Pollard1990a,Pollard1992a,Butkus12a,Cina2016,Le21a,Turner2020,Quintela2022a}. This adiabatic picture can be integrated in a number of respects, including deviations from harmonicity \cite{Park2000a,Arpin21a}, coupling between different modes \cite{Schultz22a,Yan1986a}, dependence of the vibrational frequencies on the electronic state \cite{Fidler13a}. 

The interplay between electronic and nuclear degrees of freedom is even closer in the presence of vibronic couplings, which result in coherent population transfer between the diabatic states and hopping of the vibrational wave packet between the corresponding potential energy surfaces \cite{DeSio16a}. Its effects have been observed in a variety of physical systems, ranging from molecular crystals to J-aggregates \cite{Spano2010a}, from polymeric films to natural and artificial light-harvesting systems. A detailed and quantitative explanation of the observed multidimensional spectra requires a detailed theoretical description of these complex system, and possibly of its interaction with the environment. The general understanding of the multidimensional spectra, and specifically the capability of disentangling the electronic and vibrational coherences, can be possibly favored by the investigation of relatively simple systems, such as molecular dimers \cite{Hayes13a,Halpin2014}.
On the other hand, a number of reduced models have been introduced in order to allow the rationalization of the observed spectra and to provide a semi-quantitative understanding of the underlying dynamics in terms of a few electronic levels and vibrational modes   \cite{Ishizaki10a,Tiwari13a,Butkus14a,krcmar,Duan2016,DeSio16a,Li2021a,Caycedo-Soler2022}. 

Here we consider linear and nonlinear response functions in a class of multilevel non-adiabatic model systems defined as follows. The vibrational degrees of freedom are described by harmonic oscillators, which undergo a different displacement for each of the electronic diabatic states. The Hamiltonian also includes terms that coherently couple pairs of diabatic states, thus introducing non-adiabaticity. 

The linear response functions are identified (up to a prefactor) with specific propagators, which are computed in three steps. First, the propagators are expanded in a Dyson series with respect to the non-adiabatic component of the Hamiltonian: each term in the series thus corresponds to a given number of transitions between the diabatic electronic states. For given number of transition and for given values of these transition times, the propagator can be formally (though not physically) identified with the adiabatic response functions, whose analytical expressions have been derived in Ref. \onlinecite{Quintela2022a} within a coherent state approach. After performing the Taylor expansion of such response function with respect to the relevant displacement, we integrate with respect to the interaction times, and obtain simple analytical expressions for each of the contributions. Third-order response functions are then derived, after decomposing them into the product of three propagators.  

The paper is organized as follows. In Section II we define the model systems to which the approach is applied. Section III contains the main results, namely the expressions of the single- and multiple-time propagators, and the corresponding (linear and nonlinear) response functions. Section IV contains the main steps in the formal derivation of the above results. Finally, we draw the conclusions in Section V. 

\section{The model}

\begin{figure}[h]
\centering
\includegraphics[width=0.4\textwidth]{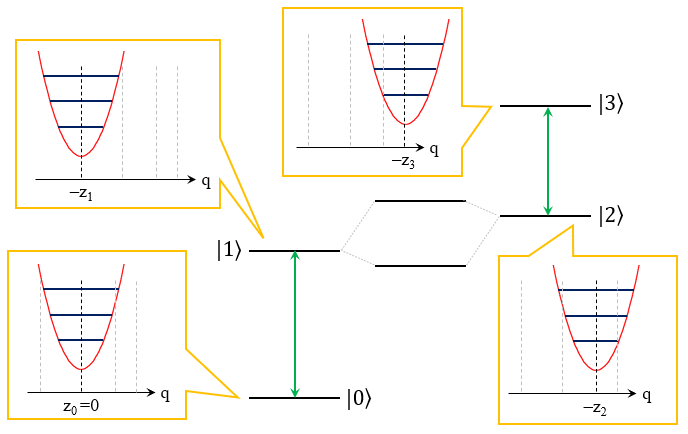}
\caption{First model system ($A$), which includes two, non-adiabatically coupled excited electronic states $|1\rangle$ and $|2\rangle$. Optical transitions (green arrows) couple $|1\rangle$ with the ground state $|0\rangle$ and $|2\rangle$ with the doubly excited state $|3\rangle$. Each electronic state $|k\rangle$ implies a displacement by $-z_k$ of the harmonic oscillator corresponding to the vibrational mode. \label{fig:1}}
\end{figure}
The present approach allows the derivation of the response function in the presence of non-adiabatic couplings between electronic and vibrational degrees of freedom. More specifically, it applies to models where the vibrational modes can be described by harmonic oscillators and the coupling between these and the electronic degree of freedom results in an electronic-state dependent displacement of the oscillators. {\color{black} The eigenstates of such displaced harmonic oscillator Hamiltonian are characterized by the factorization of the electronic and vibrational components, as results from the crude adiabatic approximation \cite{Azumi}. The non-adiaticity is introduced by a direct coupling between two electronic states, with no involvement of the vibrational degrees of freedom \cite{Witkowski}.  

Within such a class of models, we consider in the following those that are complex enough to display the processes of interest, but otherwise as simple as possible. Throughout the paper, we assume that the non-adiabatic coupling only involves the first two excited states.}
The corresponding Hamiltonian reads
\begin{align}\label{eq:ham1}
    H = H_0 + V = \sum_{\xi=0}^{N-1} H_{0,\xi} + \hbar[\eta |1\rangle\langle 2| + \eta^*|2\rangle\langle 1|] ,
\end{align}
where $V$ represents the non-adiabatic term, $H_0=\sum_{\xi=1}^N H_{0,\xi}$ includes all the adiabatic ones, and its electronic-state specific components are given by
\begin{align}\label{eq:ham2}
    H_{0,\xi} = | \xi \rangle\langle \xi | \left[\hbar\bar\omega_\xi +\sum_{\zeta=1}^{G} \hbar\omega_\zeta (a_\zeta^\dagger + z_{\zeta,\xi}) (a_\zeta+z_{\zeta,\xi})\right] .
\end{align}
In the following, and for the rest of the paper, we set $\hbar\equiv 1$.

The eigenstates of the Hamiltonian $H$ coincide with those of the adiabatic part $H_0$ for $\xi=0$ or $\xi\ge 3$. In these subspaces and for $G=1$, the eigenstates of $H$ and $H_0$ are in fact given by $|\xi; n,-z_\xi\rangle$, where $| n,-z_\xi\rangle = \mathcal{D} (-z_\xi) |n\rangle $ are the displaced Fock states. Instead, due to the non-adiabatic term $V$, the eigenstates $| \xi,-z_1\rangle$ and $| \xi,-z_2\rangle$ of $H_{0,1}$ and $H_{0,2}$ don't coincide with those of $H$, which in general don't have a simple analytical expression. 
{\color{black} Interestingly, the form of the above Hamiltonian, and specifically that of the non-adiabatic term, changes qualitatively if one replaces the basis $\{|1\rangle,|2\rangle\}$ with $\{|+\rangle,|-\rangle\}$, formed by the states that diagonalize $H_e=\sum_{\xi=1,2} \hbar\bar\omega_\xi |\xi\rangle\langle\xi|+V$. In such a basis, the coupling between electronic and vibrational degrees of freedom is has a non-diagonal component in the electronic basis, which can be identified with the non-adiabatic part of the Hamiltonian (see Appendix \ref{app:x}).}

\begin{figure}[h]
\centering
\includegraphics[width=0.4\textwidth]{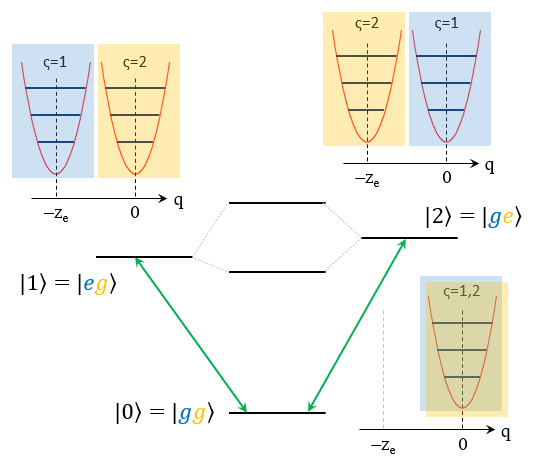}
\caption{Second model system ($B$), which includes two, non-adiabatically coupled excited electronic states $|1\rangle$ and $|2\rangle$. These correspond to the excitation respectively of the first and second monomer that form a dimer. Transitions between the ground ($g$) and excited state ($e$) of each monomer can be induced optically (green arrows). The excitation of each monomer results in a displacement by $-z_e$ of the corresponding vibrational mode (harmonic oscillator). \label{fig:2}}
\end{figure}

In the following, we refer to two simple and yet interesting model systems, corresponding to particular cases of the above Hamiltonian $H$. The first one, referred to as {\it model $A$}, is represented by a four-level system with a single vibrational mode ($N=4$, $G=1$, Fig. \ref{fig:1}). Within such model, we derive the expressions of the third-order response functions, which include contributions from processes such as excited state absorption, involving the doubly excited state $|3\rangle$. The second model, referred to as {\it model $B$}, is represented by a three-level system with two vibrational modes ($N=3$, $G=2$, Fig. \ref{fig:2}) and can be referred to a pair of coupled monomers; each monomer is coupled to its own (localized) vibrational mode. For this model we compute the first-order response function, and show how this can formally reduced to a single-mode response function in the case of a symmetric dimer. {\color{black} The dimer would in principle include a doubly excited state $|3\rangle$, which however doesn't play any role in the linear response functions considered for this model, and is thus disregarded. In fact, the present approach could in principle be applied to a single, more general model, which includes both a doubly excited state and two vibrational modes. However, this would complicate the analytical expressions and make their physical meaning less transparent, without introducing significantly new elements. For the sake of clarity, these two features are kept separate, and investigated independently from one another in the two models.}

Finally, in deriving the response functions, we assume that the system dynamics is triggered by a Franck-Condon transition between the electronic states, induced by the interaction with the external electric field. This is followed by a free evolution of the system, resulting from the interplay between the electronic and vibrational degrees of freedom. 

\section{\label{sec:main results} Main results}

\begin{figure}[h]
\centering
\includegraphics[width=0.45\textwidth]{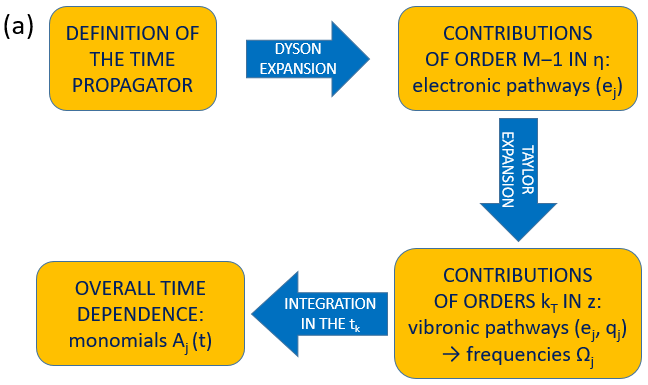}
\includegraphics[width=0.45\textwidth]{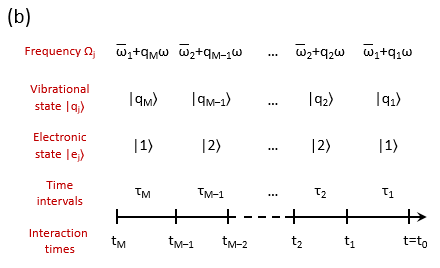}
\caption{\color{black} (a) Main steps for computing the single- and multi-time propagators: Dyson expansion with respect to the non-adiabatic coupling; Taylor expansion with respect to the displacements; integration over the interaction times. (b) Electronic and vibronic pathways (and related frequencies $\Omega_j$) associated to the terms obtained after the Dyson and Taylor expansions. The diagram refers to the case where the initial and final states concide with $|1\rangle$. \label{fig:FA1}}
\end{figure}

The central result of the present article is represented by the time propagators between the excited states belonging to the subspace $\mathcal{S}_e$. {\color{black} This, result is then used to derive the expressions of the linear and nonlinear response functions. In the following, we present a brief discursive description of the method (Subsec. \ref{subsec:1}), followed by the presentation of the final expressions (Subsecs. \ref{subsec:2}--\ref{subsec:3}). The formal derivations of the results are presented in Section IV.

\subsection{Brief description of the method\label{subsec:1}}

The relevant time propagator is the matrix element of the time evolution operator between the states $|\sigma;0\rangle$ and $|\sigma';0\rangle$: these are given by the product of the diabatic electronic states that are coupled by the non-adiabatic interaction $V$ ($\sigma,\sigma'=1,2$), and of the vibrational ground state of the undisplaced harmonic oscillator. The time-evolution operator is computed by performing a Dyson expansion with respect to $V$: each term in the expansion corresponds to an {\it electronic pathway}, {\it i.e.} to a given sequence of electronic states $e_1,\dots,e_{M}$ (an alternating sequence of $|1\rangle$ and $|2\rangle$), being $M-1$ the order of the expansion. The overall time  evolution of the system that one can associate to each electronic pathway consists of a sequence of sudden transitions between the two diabatic states, interleaved by time intervals during which the system remains in the same electronic state (Fig. \ref{fig:FA1}). 

For the vibrational state, each transition between the states $|1\rangle$ and $|2\rangle$ implies a hopping of the coherent state from one potential energy surface to the other, being these two relatively displaced parabolas. The resulting time evolution resembles that induced by sequences of delta-like laser pulses within the linearly displaced harmonic oscillator model \cite{Quintela2022a}. This formal analogy allows us to use in the present case the analytical expressions that have recently been derived for the vibrational component of the response function in the adiabatic case ($R$). 

The following step consists in the integration over all the possible values of the non-adiabatic interaction times. In order to perform such integration analytically, we perform a Taylor expansion of $R$, which can be written as the product of $M(M+1)/2$ double exponential functions. Each term in the Dyson expansion (order $M-1$ in the non-adiabatic coupling constant $\eta$) thus gives rise to a number of infinite terms, one for each set of orders $k_i$ ($i=1,\dots,M(M+1)/2$) of the Taylor expansions. Formally, each of these terms can be written as a product of exponential functions, that oscillate during the intervals of duration $\tau_j$ ($j=1,\dots,M$) with a frequency $\Omega_j$. This is given by the sum of an electronic and a vibrational contributions. The former corresponds to the energy $\bar\omega_j$ ($\hbar\equiv 1$) of the electronic state for the relevant time interval (specified by $j$) and electronic pathway (specified by $M-1$); the latter one is the energy $q_j\omega$ of the $q_j$-th eigenstate of the undisplaced harmonic oscillator. The values of $q_j$ result from those of $k_l$ in a one-to-many correspondence. Physically, one can thus associate to each term of the Taylor expansion a {\it vibronic pathway}, defined by a sequence of electronic and vibrational states $|e_j;q_j\rangle$, with $j=1,\dots,M$. Besides, each of these term is proportional to the displacements ($z_1$ or $z_2$) or their difference to the power of $k_T=\sum_{i=1}^{M(M+1)/2} k_i$. Being the modulus of the displacement typically smaller than one, this series is expected to converge, even though the number of terms increases rapidly with $k_T$. 

These functions can be analytically integrated, and give a formally simple result, consisting - for each vibronic pathway - in the sum of $M$ terms, each one oscillating at a frequency $\Omega_j$. If all these frequencies differ from one another, the oscillating terms $e^{-i\Omega_j t}$ are multiplied by constants $A_j$. If $k$ of those frequencies coincide, then each of the multiple $e^{-i\Omega_j t}$ is multiplied by a monomial $a_j t^{r_j}$, with $r_j=0,\dots,k-1$ (it follows from the calculations that the number of identical frequencies for each vibronic pathway cannot exceed $(M-1)/2$). Being this feature common to all the terms that result from the Taylor expansion, the entire contribution of order up to $M-1$ in $\eta$ is given by the sum of terms that oscillate at the frequencies $\bar\omega_1$ and $\bar\omega_2$ (diabatic state energies) and of their vibrational replicas, multiplied by polynomial functions of $t$, of order $(M-2)/2$ for even $M$ and $(M-1)/2$ for odd $M$. 

The extension of this approach to the multimode case is rather straightforward, because the dynamics of the $G$ vibrational modes are independent from one another. The Dyson expansion is of the propagator is not modified by the presence of multiple modes. On the other hand, the Taylor expansion has to be performed for each of the adiabatic response functions $R$, resulting in a larger number of vibronic pathways. Each of these is given by a sequence of states $|e_j;{\bf q}_j\rangle$, where ${\bf q}_j$ defines a $G$-dimensional vibrational (Fock) state. The final expression of the response function is thus identical to that discussed above, apart from the replacement - in the frequencies $\Omega_j$ - of the single-mode energies $q_j\omega$ with their multimode counterparts $\sum_{\zeta=1}^G \omega_\zeta q_{j,\zeta}$. 

In the multitime propagators of interest, the overall evolution of the system is divided in three time intervals ($T_L$, $T_C$, and $T_R$), delimited by optically-induced transitions between the subspace $\mathcal{S}_e$, and the ground or doubly excited states. The generalization of the above procedure thus requires two independent Dyson expansions, one for each of the time evolutions that take place in $\mathcal{S}_e$, during the waiting times $T_L$ and $T_R$ (the evolution in $T_C$ always takes place outside from the subspace $\mathcal{S}_e$, and therefore does not require a further expansion). The overall function at defined interaction times can be written as a product of $M_L(M_L+1)/2+1+M_R(M_R+1)/2$ double exponential functions, being $M_L$ ($M_R$) the order in the Dyson expansion for the first (third) time interval. In the final step, the integration is performed independently with respect to the interaction times belonging to the intervals $T_L$ and $T_R$. This gives rise to the functions of order $M_L - 1$ and $M_R - 1$ in $\eta$, and that depend respectively on $T_L$ and $T_R$, in the same way as the single-time propagators depend on $t$.
} 

\subsection{Time propagators\label{subsec:2}}

We consider the case where the system modeled by the Hamiltonian $H$ [Eqs. (\ref{eq:ham1}-\ref{eq:ham2})] undergoes a Franck-Condon transition from the ground state $|0;0\rangle$ to $|1;0\rangle$, corresponding to a generic linear superposition of Hamiltonian eigenstates. This will evolve in time, under the combined effect of the adiabatic ($H_{0,e}$) and non-adiabatic ($V$) terms.

{\color{black} More specifically, the propagators are written as the sum of different terms, each one corresponding order ($M-1$) in the non-adiabatic interactions. The nonadiabatic interactions take place at times $t_k$, with $t_{M-1} < t_{M-2} < \dots < t_1 $ and result in transitions between, e.g., states $|1; q_{k+1}\rangle$ and $|2; q_k\rangle$ (where the $q_k$ specify the Fock states of the undisplaced harmonic oscillator). Between two consecutive nonadiabatic interactions, for time intervals $\tau_k=t_{k-1}-t_k$, the system evolves freely under the effect of the Hamiltonian $H_{0}$ and accumulates the phase $\Omega_k\tau_k$. The response functions are eventually derived by integrating over the interaction times $t_k$. Between two consecutive non-adiabatic interactions, for time intervals of duration $\tau_k=t_{k-1}-t_k$, the system evolves freely, under the effect of the Hamiltonian $H_0$, and accumulates the phase $\Omega_k\tau_k$. The response functions are eventually derived by integrating over the interaction times. }

\subsubsection{Off-diagonal elements}
{\color{black}
If the number of transitions that has taken place in the time $t$ is odd, $M-1=2n+1$, the initial and final excited states differ. The resulting time propagator, {\it i.e.} the matrix element of the time-evolution operator $U_S=e^{-i H t}$, can be written in the form:
\begin{gather}
    \langle 2; 0 | U_S | 1; 0 \rangle = \sum_{n=0}^\infty \eta^* |\eta|^{2n} F_{2n+1}(t)
    \nonumber\\ = \sum_{n=0}^\infty \eta^* |\eta|^{2n} \sum_{\bf k} 
\left\{C\sum_{j=1}^{2n+2} A_{j} (t)\, e^{-i\Omega_j t}\right\}_{\bf k} ,\label{eq:a1}
\end{gather}
where it is intended that all the functions and parameters in the curly brackets depend on ${\bf k}$ (see below).
The function $F_{2n+1}(t)$, corresponding to the order $2n+1$ in the Dyson expansion, is given by the sum of monomials $A_{j}(t)=a_j t^{r_j}$, with $r_j \le n$, multiplied by terms that oscillate at the frequencies:
\begin{align}\label{eq:abc}
    \Omega_j =
    \left\{ 
    \begin{array}{c}
    \omega\, q_j + \bar\omega_1\,,\ {\rm for\ even}\ j \\
    \omega\, q_j + \bar\omega_2\,,\ {\rm for\ odd}\ j
    \end{array}\right. ,
\end{align}
with $q_j$ non-negative integers. These frequencies are thus given by the sum of two terms: the energy of the diabatic states ($\bar\omega_1$ or $\bar\omega_2$), and an integer multiple of the vibrational frequency $\omega$.

Each of the $(2n+1)$-th order terms in the Dyson expansion [Eq. \ref{eq:abc}] is given by the sum of different contributions, one for each vector ${\bf k}=(k_1,\dots.k_{M(M+1)/2})$. These contributions result from the Taylor expansion of the adiabatic propagator, and are of order $k_T=\sum_{i=1}^{M(M+1)/2} k_i$ in the displacements $z_{\zeta,\xi}$ [see Eq. (\ref{eq:ham2})]. The explicit dependence of the contributions in the sum on ${\bf k}$ and on the displacements can be expressed as follows:
\begin{gather}
    \langle 2; 0 | U_S | 1; 0 \rangle = \sum_{n=0}^\infty \eta^* |\eta|^{2n}  \nonumber\\ \sum_{{\bf k}} \left\{\left[(-i)^{2n+1} e^{h_{M}({\bf z})} \chi_{M} ({\bf z},{\bf k}) \right]  \sum_{j=1}^{2n+2} A_j(t)\, e^{-i\Omega_j t} \right\}_{\bf k}.
\end{gather}
where $M=2n+1$.

The frequencies $\Omega_j$ and the functions $A_j(t)$ depend on ${\bf k}$ only through the integers $q_j$, which specify the sequence of vibrational states in the related pathway. These integers are given by the expression:
\begin{gather}
    q_j = \sum_{x=1}^M \ \sum_{y=\max(1,j-x+1)}^{\min(j,M+1-x)} k_{(x-1)M-\frac{1}{2}(x-1)(x-2)+y}.
\end{gather}
We note that the relation between ${\bf k}$ and ${\bf q}$ is not one-to-one, for different vectors ${\bf k}$ can correspond to a same ${\bf q}$.

The constant prefactor, denoted with $C$ in Eq. (\ref{eq:abc}), depends both on ${\bf k}$ and on the vector ${\bf z}=({\rm z}_1,\dots,{\rm z}_M)$, whose components coincide with the displacements of the oscillator (here, these are given by ${\rm z}_k=z_2$ for odd $k$ and ${\rm z}_k=z_1$ for even $k$). 
Such dependence is expressed by the functions $h_M$ and $\chi_M$.}
The former one, whose general expression is reported in Section \ref{sec:derivations}, is here given by 
\begin{gather}
    h_M ({\bf z}) = -\frac{1}{2} M z_{12}^2 - z_1 z_2,
\end{gather}
where $z_{ij}\equiv z_i-z_j$.
The latter one $\chi_M$, which depends both on ${\bf z}$ and on ${\bf k}$, in the present case reads
\begin{gather}
    \chi_M ({\bf k},{\bf z}) = \frac{\prod_{p=1}^{M-1} 
    [(-1)^{p} z_1 z_{21}]^{k_{1+w}}
    [(-1)^{p} z_2 z_{12}]^{k_{M-p+1+w}}}{\prod_{l=1}^{M(M+1)/2} k_l!} \nonumber\\
    \times (z_1 z_2)^{k_{M(M+1)/2}} \prod_{q=2}^{M-p}[(-1)^{p+1} z_{12}^2]^{k_{q+w}},
    \label{eq:s1}
\end{gather}
where $w=(p-1)M-(p-1)(p-2)/2$. {\color{black} The zero-phonon line corresponds to ${\bf q}={\bf 0}$ and $\chi_M=1$.}

The expansion in Eq. (\ref{eq:a1}) includes in principle an infinite number of terms, {\color{black} resulting from both the Dyson and the Taylor expansions. However the relative importance in the former expansion is expected to decrease for increasing values of the order $2n+1$, especially in the short-time limit ($|\eta| t \lesssim 1$). As to the second expansion, being in general $|z_1|,|z_2|<1$, the value of the constant prefactor $C$ is also expected to rapidly decrease for increasing values of the order $k_T$, which defines the power in the displacements.}


\subsubsection{Diagonal elements}

If the number of transitions that has taken place in the time $t$ is even, the initial and final excited states coincide. The resulting propagators read:
{\color{black}
\begin{gather}
    \langle \sigma; 0 | U_S | \sigma; 0 \rangle = \sum_{n=0}^\infty |\eta|^{2n} F_{2n}(t)
    \nonumber\\ = \sum_{n=0}^\infty |\eta|^{2n} \sum_{\bf k} 
\left\{C\sum_{j=1}^{2n+1} A_{j} (t)\, e^{-i\Omega_j t}\right\}_{\bf k} , \label{eq:a2}
\end{gather}
where $\sigma=1,2$. In the $0$-th order contribution ($n=0$), the propagator is reduced to that derived for the adiabatic case \cite{Quintela2022a}. Analogously to the case of the off-diagonal elements, the functions $F_{2n}$ are given by the sum of monomial functions $A_j(t) = a_j t^{n_j}$ (with $n_j \le n$), multiplied by terms that oscillate at the frequencies
\begin{align}\label{eq:abc}
    \Omega_j =
    \left\{ 
    \begin{array}{c}
    \omega\, q_j + \bar\omega_{3-\sigma}\,,\ {\rm for\ even}\ j \\
    \omega\, q_j + \bar\omega_\sigma\,,\ {\rm for\ odd}\ j
    \end{array}\right. .
\end{align}

As to the dependence of the different contributions on ${\bf k}$, resulting from the Taylor expansion, this is given by:
\begin{gather}
    \langle \sigma; 0 | U_S | \sigma; 0 \rangle = \sum_{n=0}^\infty |\eta|^{2n}  \nonumber\\ \sum_{{\bf k}} \left\{\left[(-i)^{2n} e^{h_{M}({\bf z})} \chi_{M} ({\bf z},{\bf k}) \right]  \sum_{j=1}^{2n+2} A_j(t)\, e^{-i\Omega_j t} \right\}_{\bf k},
\end{gather}
where $M=2n+1$.
}
The functions $h_M$ and $\chi_M$ take here different forms with respect to the previous case. In fact, the function $h_M$ of the displacements is given by 
\begin{gather}\label{eq:h1}
    h_M ({\bf z}) = -\frac{1}{2} (M-1) z_{12}^2 - z_\sigma^2,
\end{gather}
where the vector ${\bf z}$ has components ${\rm z}_k=z_\sigma$ for odd $k$ and ${\rm z}_k=z_{3-\sigma}$, for even $k$.
The function $\chi_M$, which depends both on ${\bf z}$ and on ${\bf k}$, reads
\begin{gather}
    \chi_M ({\bf k},{\bf z}) = \frac{\prod_{p=1}^{M-1} 
    [(-1)^{p} z_\sigma z_{3-\sigma,\sigma}]^{k_{1+w}+k_{M-p+1+w}}}{\prod_{l=1}^{M(M+1)/2} k_l!} \nonumber\\ \label{eq:chi1}
    \times z_\sigma^{2k_{M(M+1)/2}} \prod_{q=2}^{M-p}[(-1)^{p+1} z_{12}^2]^{k_{q+w}},
\end{gather}
where $w=(p-1)M-(p-1)(p-2)/2$. {\color{black} The zero-phonon line corresponds to ${\bf q}={\bf 0}$ and $\chi_M=1$.} 

\begin{figure}[h]
\centering
\includegraphics[width=0.23\textwidth]{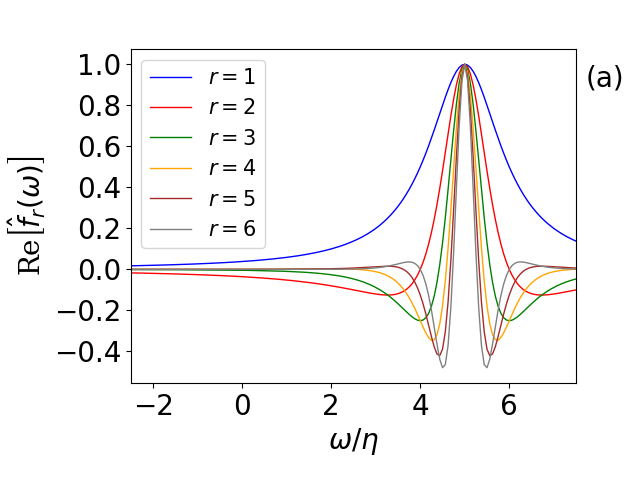}
\includegraphics[width=0.23\textwidth]{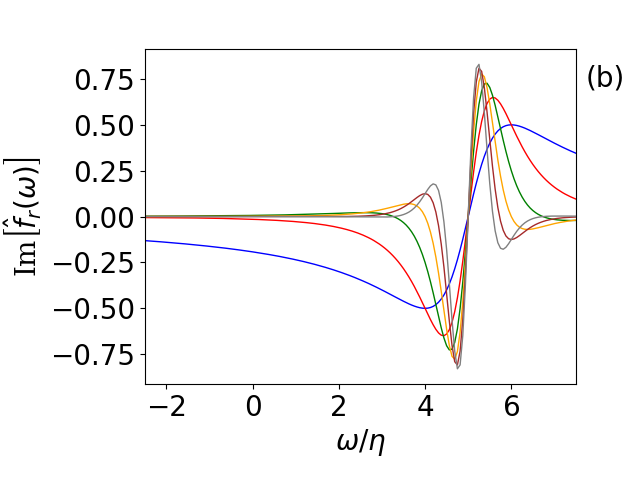}
\includegraphics[width=0.23\textwidth]{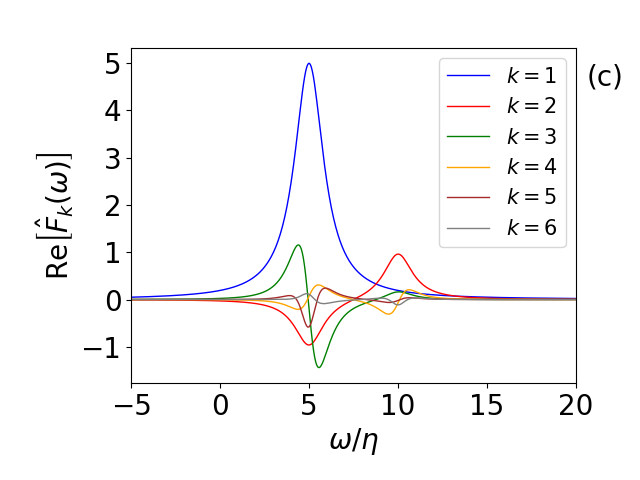}
\includegraphics[width=0.23\textwidth]{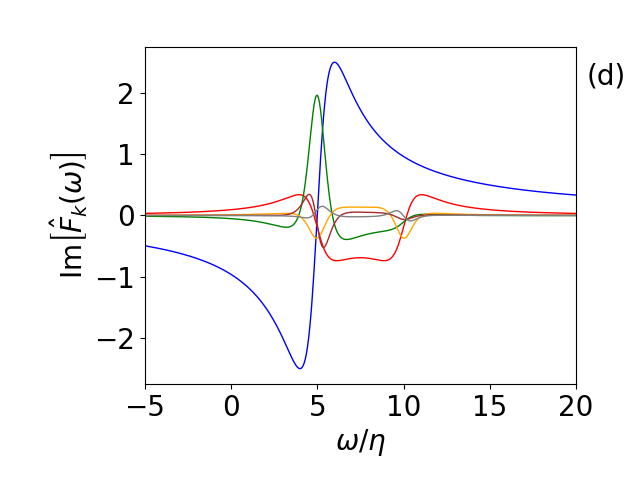}
\includegraphics[width=0.23\textwidth]{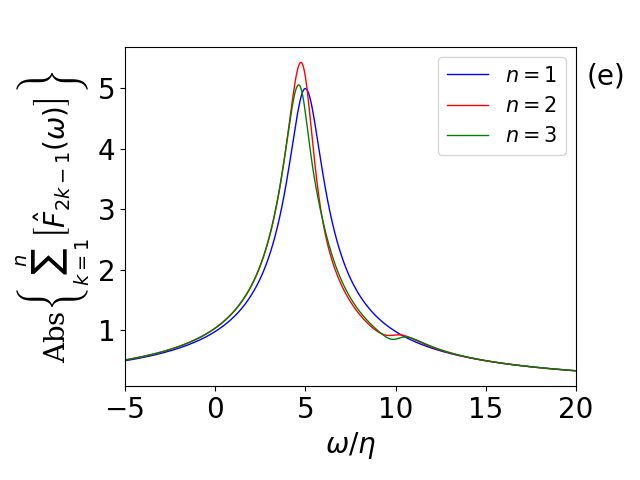}
\includegraphics[width=0.23\textwidth]{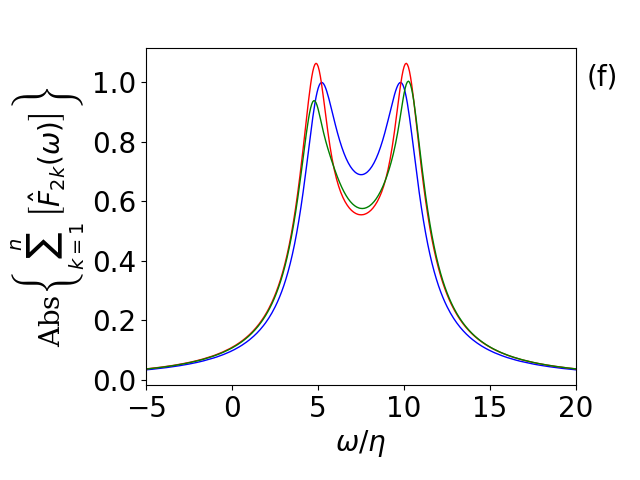}
\caption{\color{black} Real and imaginary parts of the Fourier transforms of: (a,b) the functions $f_r(t)=t^r e^{-i\bar\omega_1 t}$; (c,d) contributions to the propagators. Absolute values of the diagonal (e) and off-diagonal (f) propagators, up to different orders in the Dyson expansion. In all cases, we consider the following parameter values: $z_1 = z_2 = 0$, $\bar\omega_1= 5$, $\bar\omega_2 = 10$, and $\gamma = 1$, all in units of $\eta$, which is assumed to be real and positive.\label{fig:FA2}}
\end{figure}

{\color{black}
\subsection{Propagator in the frequency domain \label{subsec:Pitsd}}

The propagators are shown to consist of a number of contributions, whose time dependence is given by functions $f_r(t)=A(t)\,e^{-(\gamma+i\Omega) t}$, where $A(t)=a t^{r}$ and the exponential decay ($\gamma >0$) results from decoherence (see Subsec. \ref{subsec:deco}). Therefore, the Fourier transform of the propagator is given by combinations, with equal coefficients, of the $\hat{f}_r(\omega) = {\rm FT} \{f_r(t)\}$, whose expressions read:
\begin{align}
    \hat{f}_r(\omega) \!=\! a \!\int^{\infty}_0\! dt \,  t^{r} \, e^{-[\gamma+i(\Omega-\omega)] t} \!=\! \frac{r!}{[\gamma+i(\Omega-\omega)]^{r+1}} .
\end{align}
In Fig. \ref{fig:FA2}(a,b) we plot the real and imaginary parts of the functions $\hat{f}_r(\omega)$ corresponding to different values of $r$. 

Due to the complex character of the prefactors that appear in the expression of the functions $A_j(t)$ (see Appendix \ref{app:lof}), both the real and the imaginary part of each contribution in the Dyson expansion of the propagator's Fourier transform
\begin{gather}
\langle \sigma; 0 | \hat{U}_S (\omega) | \sigma'; 0 \rangle = {\rm FT} \{\langle \sigma; 0 | U_S (t) | \sigma'; 0 \rangle\} 
\end{gather}
consist of combinations of real and imaginary parts of the functions $\hat{f}_r(\omega)$ [panels (c,d)], and thus present a mixed absorptive and dispersive character. 

We finally apply these results to the diagonal and off-diagonal propagators, up to different orders in the Dyson expansion. For the sake of simplicity, we show this in the case of the undisplaced oscillator ($z_1=z_2=0$), where $h_M=0$, $\chi_M=1$, $\Omega_{2k-1}=\bar\omega_\sigma$, and $\Omega_{2k}=\bar\omega_{3-\sigma}$ (being $q_j=0$ for all the $j$). The propagators are given by the sum of two terms that oscillate at the diabatic state energies, $e^{-i\bar\omega_1 t}$ and $e^{-i\bar\omega_2 t}$, each one multiplied by a polynomial of order $n$ (the expressions of the monomials $A_j(t)$, in general and specifically for the case of the undisplaced oscillator, are given in Appendix \ref{app:lof} for $M \le 6$). 

In this particular case,  the diagonal propagator is dominated by the zero-th order contribution ($M=1$) in the diagonal case (peak at $\bar\omega_1$), corresponding to a diabatic evolution within the initial state $|1\rangle$, with a minor contribution at $\bar\omega_2$, resulting mainly from the second-order term ($|\eta|^2$, transitions $|1\rangle\longrightarrow |2\rangle\longrightarrow |1\rangle |$)  
[panel (e)]. The off-diagonal propagator presents two symmetric peaks at the two frequencies $\bar\omega_1$ and $\bar\omega_2$, mainly resulting from first-order contribution ($M=2$), and corresponding to the occurrence of a single non-adiabatic transition $|1\rangle\longrightarrow |2\rangle$.

From the expressions of the Fourier transforms, it follows that the relative weight of the contributions corresponding to different orders is given by the values of the diabatic gap $\bar\omega_{12}$ and of the relevant decay rate $\gamma$, relative to the non-adiabatic coupling $\eta$.
In fact, the terms of order $M-1$ and resulting from a monomial $A_j(t)$ of order $r$ are proportional (at resonance) to 
\begin{gather}
    \frac{|\eta|^{M-1}}{|\bar\omega_{12}|^{M-r-1} \gamma^{r}}.
\end{gather}
The smaller $\gamma$, the larger the relative weight of the terms with high $r$.
The convergence (the fact that the contributions lose weight for increasing $M$) results from the condition $|\eta| < |\bar\omega_{12}|,\gamma$. }

\subsection{Linear response function}

\begin{figure}[h]
\centering
\includegraphics[width=0.5\textwidth]{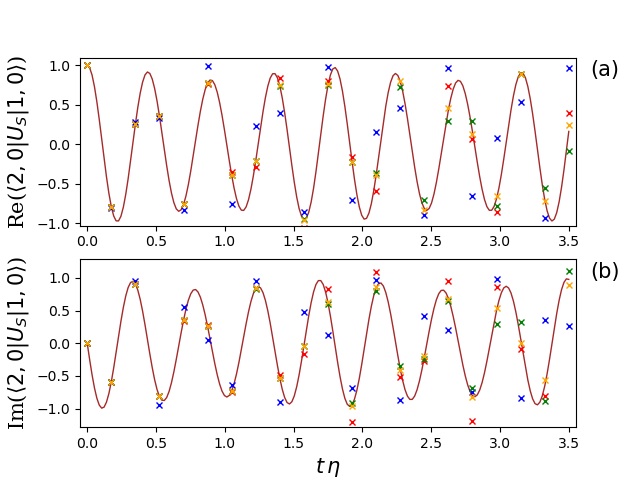}
\caption{\label{fig:3} Real (a) and imaginary (b) parts of the propagator corresponding to model $B$, obtained by including the non-adiabatic interaction up to different orders $2n$: 0 (blue symbols), 2 (red), 4 (green), 6 (orange). The solid curve corresponds to the propagator obtained by an independent approach, based on the diagonalization of the full Hamiltonian. The Hamiltonian parameters are: $z_1=0.1$ and $z_2=0$ for the first mode, $z_1=0$ and $z_2=0.1$ for the second mode; the frequencies are $\omega=1.587$, $\bar\omega_1=14.29$ and $\bar\omega_2=17.14$, all in units of $\eta$; the sum on the vectors ${\bf k}$ includes all terms with $k_T=\sum_i k_i \le 8$. {\color{black} The time is given in units of $1/\eta$, being $\eta$ the non-adiabatic coupling, assumed to be real and positive.}}
\end{figure}
A first, straightforward application of the propagators reported in the previous Subsection is represented by the first-order response function, for the model systems $A$ and $B$, schematized respectively in Fig. \ref{fig:1} and Fig. \ref{fig:2}. 

\paragraph{\color{black} Model $A$.}
Model $A$ is characterized by the presence of one vibrational mode, and only one excited state that can be optically addressed from the ground state. The resulting response function is given by:  
\begin{gather}\label{eqx02}
    \mathcal{R}^{(1)}_A (T_1)
    =i|\mu_{01}|^2 \sum_{n=0}^\infty (-i)^{2n} |\eta|^{2n} e^{h_{M}}
    \sum_{{\bf k}} \chi_{M} 
    f_{M,{\bf q},1}(T_1) ,
\end{gather}
where $h_{M}({\bf z})$ and $\chi_{M}({\bf k},{\bf z})$ are given respectively by Eq. (\ref{eq:h1}) and Eq. (\ref{eq:chi1}), while the time dependence is given by
\begin{gather}
    f_{M,{\bf q},\sigma}(T_1) = \sum_{j=1}^M A_j(T_1)\, e^{i\Omega_j T_1} .
\end{gather}
In the response function, only even-order contributions in the non-adiabatic interaction matter, because also the emission process at the end of the time evolution has to take place from the excited state $|1\rangle$. Therefore, $M=2n+1$ and the vector ${\bf z}$ has components ${\rm z}_k=z_\sigma$ for odd $k$, and ${\rm z}_k=z_{3-\sigma}$ for even $k$. 

{\color{black} 

In the presence of relaxation and dephasing (Subsec. \ref{subsec:deco}), the above response function undergoes an exponential decay as a function of $T_1$. In particular, this results in a prefactor
\begin{gather}
    F_A(T_1) = e^{-(\gamma_g+\gamma_e+\Gamma_e/2)T_1} 
\end{gather}
to be added to the above expression of $\mathcal{R}^{(1)}_A (T_1)$.}

\paragraph{\color{black} Model B.}
The case of model $B$ is conceptually equivalent to the previous one, but includes some additional contribution. This is due to the presence of a second vibrational mode and of a second allowed optical transition, that between the states $|0\rangle$ and $|2\rangle$. As a result, in the case of a symmetric dimer ($\bar\omega_1\!=\!\bar\omega_2\!\equiv\!\omega_e$, $\mu_{0,1}\!=\!\mu_{0,2}\!\equiv\!\mu_e$, $z_{1,1}\!=\!z_{2,2}\!\equiv\! z_e$), the linear response function reads:
\begin{gather}
    \mathcal{R}^{(1)}_B (T_1) = i|\mu_{0e}|^2 \left[
    \sum_{n=0}^\infty (-i)^{2n+1}(\eta +\eta^*) |\eta|^{2n} e^{-(2n+2)z_{e}^2} \right. \nonumber\\ \sum_{{\bf k}} \chi_{2n+2}' \,  
    f_{2n+2,{\bf q}}(T_1) +
    2 \sum_{n=0}^\infty (-i)^{2n} |\eta|^{2n} e^{-(2n+1)z_{e}^2} \nonumber\\ \left. \sum_{{\bf k}} \chi_{2n+1}' \,  
    f_{2n+1,{\bf q}}(T_1)\right] .
\end{gather}
Here, the first (second) term in square brackets corresponds to pathways with an odd (even) number of non-adiabatic processes, such that the absorption and emission processes involve different (the same) excited states. We note that, due to the degeneracy between the two excited states, $\bar\omega_{12}=0$ and the $f_{M,{\bf q},1}=f_{M,{\bf q},2}\equiv f_{M,{\bf q}}$.

The fact that the two-level systems are identical implies that the vibrational modes are characterized by the same frequency and undergo the same displacement $z_e$ in passing from the ground state $|g\rangle$ to the excited state $|e\rangle$. This leads to a simplification of the propagator and of the resulting response function, which can be written formally as in the single-mode case, apart from the replacement of $h_M ({\bf z})$ with $h_M' ({\bf z}) = - M z_{e}^2 $ and of the function $\chi$ with
\begin{gather}
    \chi_M' ({\bf k},z_e) = \frac{\prod_{p=1}^{M-1} [(-1)^{p+1} z^2_e]^{k_{1+w}+k_{M-p+1+w}}}{\prod_{l=1}^{M(M+1)/2} k_l!} \nonumber\\
    \times \frac{1}{2} [1-(-1)^M] z_e^{2k_{M(M+1)/2}} \prod_{q=2}^{M-p}[2(-1)^{p+1} z_{12}^2]^{k_{q+w}},
\end{gather}
where $w=(p-1)M-(p-1)(p-2)/2$.

{\color{black} In the presence of dephasing and decoherence, the response function decays exponentially as a function of time. Such decay is described by the prefactor $F_B(T_1)=F_A(T_1)$.}

\paragraph{\color{black} Verification against numerical results.}
In order to test the approach, we compare the response function obtained with the present approach with one computed with a completely independent method. This consists in diagonalizing $H$ and propagating the initial state $|1;0\rangle$ by expanding it in the basis of the Hamiltonian eigenstates. As shown in Fig. \ref{fig:3}, the results of the perturbative approach (symbols) converge to the nonperturbative results (solid line) for increasing number of terms in the expansion. Terms of increasing order are clearly required for increasing time $t$. In this particular case, a good agreement for $t |\eta| > 1$ requires the inclusion of terms up to 6-th order in the non-adiabatic coupling $V$. {\color{black} In general, from the expression of the functions $A_j(t)$ (see Appendix \ref{app:lof}) it follows that the expansion should converge for small values of $|\eta/\bar{\omega}_{12}|$ and of $t|\eta|$ ($\hbar\equiv 1$).}

\subsection{Nonlinear response function\label{subsec:3}}

The expression of the single-time propagator represents a starting point for the derivation of multi-time propagators, which can be directly related to nonlinear response functions. In particular, we focus hereafter on the response functions of third order in the light-matter interaction for model $A$ (Fig. \ref{fig:1}).  

Third-order response functions are expressed with respect to the waiting times $T_1$, $T_2$, and $T_3$, corresponding to the time intervals between consecutive interactions with the field. Besides, one can distinguish between the different contributions (pathways), based on the underlying physical process: ground-state bleaching, stimulated emission, photo-induced absorption, and double quantum coherence. In the following, the two inequivalent contributions to the response functions are derived for each of these processes. The functions $h_M$ and $\chi_M$ of the displacements are however common to all the cases, and are reported hereafter.
The function $h_M$ is given by 
\begin{gather}\label{eq:hM}
    h_M ({\bf z}) = -\sum_{p=1}^M \sum_{q=1}^{M-p+1} z_{j_{q-1} j_{q}} z_{j_{q+p-1} j_{q+p}},
\end{gather}
where $z_{ij}\equiv z_i-z_j$.
The function $\chi_M$, which also depends on ${\bf k}$, reads
\begin{gather}\label{eq:chiM}
    \chi_M ({\bf k},{\bf z}) = \frac{\prod_{p=1}^M \prod_{q=1}^{M-p+1} (z_{j_{q-1} j_{q}} z_{j_{q+p-1} j_{q+p}})^{k_{q+w}}}{\prod_{l=1}^{M(M+1)/2} k_l!},
\end{gather}
where $w=(p-1)M-(p-1)(p-2)/2$.

\subsubsection{Ground-state bleaching}

\begin{figure}[h]
\centering
\includegraphics[width=0.5\textwidth]{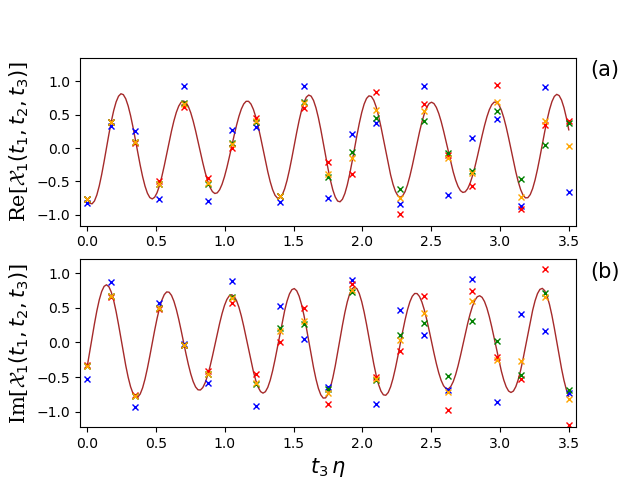}
\caption{\label{fig:4} Real (a) and imaginary (b) parts of the multitime propagator for model $A$ and corresponding, up to a constant prefactor, to the third-order response function: ground state bleaching, rephasing contribution. 
The symbols correspond to the results obtained with the perturbative approach, by including terms up to a given order $2(n_L+n_R)$ in the non-adiabatic coupling $V$: 0 (blue), 2 (red), 4 (green), 6 (orange). The solid line represents the results obtained by an independent nonperturbative approach.
The Hamiltonian parameters are: $z_1=0.1$, $z_2=0.2$ and $z_3=0.15$; the frequencies are $\omega=1.587$, $\bar\omega_1=14.29$ and $\bar\omega_2=17.14$, all in units of $\eta$; the sum on the vectors ${\bf k}$ includes all terms with $k_T=\sum_i k_i \le 8$. The plots report the dependence on $T_3$, for $T_1 = 0.7$ and $T_2=0$. {\color{black} The times are given in units of $1/\eta$, being $\eta$ the non-adiabatic coupling, assumed to be real and positive.}}
\end{figure}
The ground state bleaching is associated with those pathways where both the ket and the bra are in the ground state during the second waiting time. It includes a rephasing and a non-rephasing contribution, which are treated separately hereafter. 

\paragraph{\color{black} Rephasing contribution.}
The rephasing contribution corresponds in the perturbative (or Mukamelian) approach to the following sequence of transitions between operators: 
$| 0 \rangle\langle 0 | \longrightarrow 
 | 0 \rangle\langle j | \longrightarrow 
 | 0 \rangle\langle 0 | \longrightarrow 
 | k \rangle\langle 0 | \longrightarrow 
 | 0 \rangle\langle 0 | $,
where $|j\rangle$ and $|k\rangle$ are optically excited states. 
In the case of model $A$, one has that $j=k=1$. 
The response function reads:
\begin{gather}
    \mathcal{R}^{(3)}_2 
    =-i|\mu_{01}|^4\sum_{n_L,n_R=0}^\infty (-i|\eta|)^{2(n_L+n_R)} e^{h_M}
    \nonumber\\  \sum_{{\bf k}} \chi_{M} \, 
    f_{M_L,{\bf q}_L,1} (-T_1)\, f_{M_R,{\bf q}_R,1} (T_3)\, 
    e^{i m_C \omega (T_2+T_3)} .
\end{gather}
The $M$-dimensional vector ${\bf z}$ has the $l$-th component ${\rm z}_l=z_{j_l}$, where all the odd-numbered indices are $j_{2k+1}=1$ and all the even-numbered are $j_{2k}=2$, apart from $j_0=j_{M_L+1}=j_{M+1}=0$. The overall order $M-3=2(n_L+n_R)$ in the non-adiabatic coupling results from $2n_L$ ($2n_R$) virtual transitions in the evolution of the bra (ket) during the first (third) waiting time. 

{\color{black} In the presence of decoherence (Subsec. \ref{subsec:deco}), the above response function is multiplied by a factor $F_2$, which is given by the following expression:
\begin{gather}
    F_2 (T_1,T_2,T_3) = e^{-(\gamma_e+\gamma_g+\Gamma_e/2)(T_1+T_3)}.
\end{gather}
This accounts for the decay of the coherence between the ground state $|0\rangle$ and an arbitrary linear superposition of the states $|1\rangle$ and $|2\rangle$ that takes place during the first and third waiting times, and for the relaxation of the excited states. }

\paragraph{\color{black} Non-rephasing contribution.}
The non-rephasing contribution corresponds to the following sequence of transitions: 
$| 0 \rangle\langle 0 | \longrightarrow 
 | k \rangle\langle 0 | \longrightarrow 
 | 0 \rangle\langle 0 | \longrightarrow 
 | j \rangle\langle 0 | \longrightarrow 
 | 0 \rangle\langle 0 | $,
 where $|j\rangle$ and $|k\rangle$ are optically excited states. 
In the case of model $A$, one has that $j=k=1$. 
The expression of this contribution reads:
\begin{gather}
    \mathcal{R}^{(3)}_5
    =-i|\mu_{01}|^4\sum_{n_L,n_R=0}^\infty (-i|\eta|)^{2(n_L+n_R)} e^{h_M}  \nonumber\\ \sum_{{\bf k}} \chi_M \, 
    f_{M_L,{\bf q}_L,1} (T_3)\, f_{M_R,{\bf q}_R,1} (T_1)\, 
    e^{-i m_C \omega T_2} .
\end{gather}
The $M$-dimensional vector ${\bf z}$ has the $l$-th component ${\rm z}_l=z_{j_l}$, where all the odd-numbered indices are $j_{2k+1}=1$ and all the even-numbered are $j_{2k}=2$, apart from $j_0=j_{M_L+1}=j_{M+1}=0$. The overall order $M-3=2(n_L+n_R)$ in the non-adiabatic coupling results from $2n_L$ ($2n_R$) virtual transitions in the evolution of the ket during the third (first) waiting time.

{\color{black} Decoherence affects the non-rephasing contribution in the same way as the rephasing one. Correspondingly, the above response function has to be multiplied by a factor $F_5(T_1,T_2,T_3)=F_2(T_1,T_2,T_3)$.}

\paragraph{\color{black} Verification against numerical results.}
In order to test these analytical results, we compare the third-order response function obtained for the rephasing contribution with that derived by numerical diagonalization of the Hamiltonian. As shown in Fig. \ref{fig:4}, the results of the perturbative approach (symbols) converge to the nonperturbative results (solid line) for increasing number of terms in the expansion. Terms of increasing order are clearly required for increasing values of $T_3$ and (not shown) of $T_1$. The value of $T_2$ is irrelevant in this perspective, because non-adiabatic transitions can take place during the second waiting time, when the system state evolves within the ground state manifold.

\subsubsection{Stimulated emission}

The stimulated emission is associated with those paths where both  the  ket  and  the  bra  are  in  the  excited-state subspace $\mathcal{S}_e$  state  during  the  second  waiting  time.  It  includes  a  rephasing  and  a  non-rephasing contribution. 

\paragraph{\color{black} Rephasing contribution.}
The rephasing contribution corresponds to transitions 
$| 0 \rangle\langle 0 | \longrightarrow 
 | k \rangle\langle 0 | \longrightarrow 
 | k \rangle\langle j | \longrightarrow 
 | 0 \rangle\langle j | \longrightarrow 
 | 0 \rangle\langle 0 | $,
 where $|j\rangle$ and $|k\rangle$ are optically excited states, here (model $A$) coinciding with $|1\rangle$. Its expression reads:
\begin{gather}
    \mathcal{R}^{(3)}_1
    =-i |\mu_{01}|^4 \sum_{n_L,n_R=0}^\infty (-i|\eta|)^{2(n_L+n_R)} 
    e^{h_M} \nonumber\\ \sum_{{\bf k}} \chi_M \, 
    f_{M_L,{\bf q}_L,1} (-T_1)\, f_{M_R,{\bf q}_R,1} (T_2+T_3)\, 
    e^{i m_C \omega  T_3}.
\end{gather}
The $M$-dimensional vector ${\bf z}$ has the $l$-th component ${\rm z}_l=z_{j_l}$, where all the odd-numbered indices are $j_{2k+1}=1$ and all the even-numbered are $j_{2k}=2$, apart from $j_0=j_{M_L+1}=j_{M+1}=0$. The overall order $M-3=2(n_L+n_R)$ in the non-adiabatic coupling results from $2n_L$ ($2n_R$) virtual transitions in the evolution of the bra (ket) during the first (second and third) waiting time(s). 

{\color{black} In the presence of decoherence (Subsec. \ref{subsec:deco}), the above response function is multiplied by a factor $F_1$, which is given by the following expression:
\begin{gather}
    F_1 (T_1,T_2,T_3) = e^{-(\gamma_e+\gamma_g+\Gamma_e/2)(T_1+T_3)-\Gamma_eT_2}.
\end{gather}
This accounts not only for the dephasing and relaxation processes that affect the coherences during the waiting times $T_1$ and $T_3$ (as for the contributions related to ground state bleaching), but also for the relaxation taking place during the second waiting time $T_2$. }

\paragraph{\color{black} Non-rephasing contribution.}
The non-rephasing contribution corresponds to transitions 
$| 0 \rangle\langle 0 | \longrightarrow 
 | j \rangle\langle 0 | \longrightarrow 
 | j \rangle\langle k | \longrightarrow 
 | j \rangle\langle 0 | \longrightarrow 
 | 0 \rangle\langle 0 | $,
 where $|j\rangle$ and $|k\rangle$ are optically excited states, here coinciding with $|1\rangle$ (model $A$). Its expression reads:
\begin{gather}
    \mathcal{R}^{(3)}_4
    =-i |\mu_{01}|^4 \sum_{n_L,n_R=0}^\infty (-i|\eta|)^{2(n_L+n_R)} 
    e^{h_M} \sum_{{\bf k}} \chi_M \nonumber\\ 
    f_{M_L,{\bf q}_L,1} (-T_2)\, f_{M_R,{\bf q}_R,1} (T_1\!+\!T_2\!+\!T_3)\, 
    e^{i m_C \omega T_3} .
\end{gather}
The $M$-dimensional vector ${\bf z}$ has the $l$-th component ${\rm z}_l=z_{j_l}$, where all the odd-numbered indices are $j_{2k+1}=1$ and all the even-numbered are $j_{2k}=2$, apart from $j_0=j_{M_L+1}=j_{M+1}=0$. The overall order $M-3=2(n_L+n_R)$ in the non-adiabatic coupling results from $2n_L$ ($2n_R$) virtual transitions in the evolution of the bra (ket) during the second (three) waiting time(s). 

{\color{black} The effect of decoherence on the non-rephasing contribution coincides with that on the rephasing one. Therefore, the above response function has to be multiplied by a factor $F_4(T_1,T_2,T_3)=F_1(T_1,T_2,T_3)$.}

\subsubsection{Excited state absorption}

\begin{figure}[h]
\centering
\includegraphics[width=0.5\textwidth]{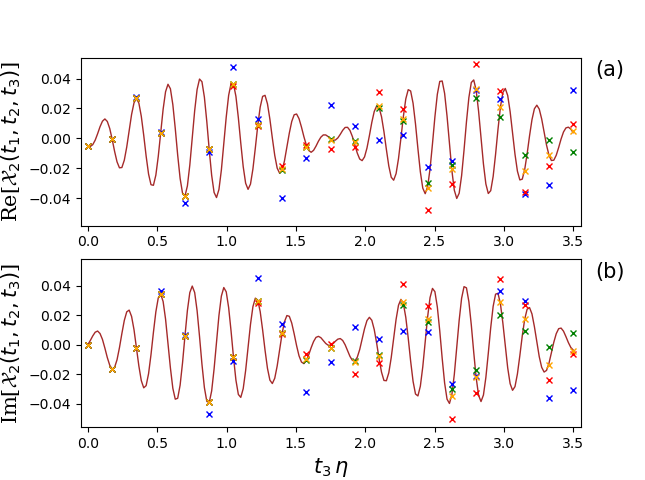}
\caption{\label{fig:5} Real (a) and imaginary (b) parts of the of the multitime propagator for model $A$ and corresponding, up to a constant prefactor, to the third-order response function: excited state absorption, rephasing contribution. 
The symbols correspond to the results obtained with the perturbative approach, by including terms up to a given order $2(n_L+n_R)$ in the non-adiabatic coupling $V$: 2 (blue), 4 (red), 6 (green), 8 (orange). The solid line represents the results obtained by an independent nonperturbative approach.
The Hamiltonian parameters are: $z_1=0.1$, $z_2=0.2$ and $z_3=0.15$; the frequencies are $\omega=1.587$, $\bar\omega_1=14.29$ and $\bar\omega_2=17.14$, all in units of $\eta$; the sum on the vectors ${\bf k}$ includes all terms with $k_T=\sum_i k_i \le 8$. The plots report the dependence on $T_3$, for $T_1= 0$ and $T_2=0.07$. {\color{black} The times are given in units of $1/\eta$, being $\eta$ the non-adiabatic coupling, assumed to be real and positive.}}
\end{figure}
The  excited  state  absorption  is  associated  to  those paths  where  both  the  ket  and  the  bra  are  in  an  excited  state subspace $\mathcal{S}_e$  during  the  second  waiting  time,  and  the ket  undergoes  a  further  excitation  process  at  the  end of  such  period.  

\paragraph{\color{black} Rephasing contribution.}
The response function associated to the rephasing contribution corresponds to transitions 
$| 0 \rangle\langle 0 | \longrightarrow 
 | 0 \rangle\langle j | \longrightarrow 
 | k \rangle\langle j | \longrightarrow 
 | l \rangle\langle j | \longrightarrow 
 | j \rangle\langle j | $,
 where $|j\rangle$ and $|k\rangle$ are singly excited states, while $|l\rangle$ is doubly excited. In the case of model $A$, one has that $j=k=1$ and $l=3$. 
The expression of the response function reads:
\begin{gather}
    \mathcal{R}^{(3)}_3
    =i |\mu_{01}\mu_{23}|^2 \sum_{n_L,n_R=0}^\infty (-i|\eta|)^{2(n_L+n_R+1)}
    e^{h_M} \sum_{{\bf k}} \chi_M \nonumber\\ \, 
    f_{M_L,{\bf q}_L,1} (\!-T_1\!-\!T_2\!-\!T_3)\, f_{M_R,{\bf q}_R,2} (T_2)\, 
    e^{-i (m_C \omega +\omega_3) T_3} .
\end{gather}
The $M$-dimensional vector ${\bf z}$ has the $l$-th component ${\rm z}_l=z_{j_l}$, where all the odd-numbered indices are $j_{2k+1}=1$, apart from $j_{M_L+1}=3$, and all the even-numbered are $j_{2k}=2$, apart from $j_0=j_{M+1}=0$. The overall order $M-3=2(n_L+n_R)+2$ in the non-adiabatic coupling results from $2n_R+1$ ($2n_L+1$) virtual transitions in the evolution of the ket (bra) during the second (three) waiting time(s). 

{\color{black} Decoherence affects the above response function (Subsec. \ref{subsec:deco}). Its effect can be accounted by including a factor $F_3$, which reads:
\begin{gather}
    F_3 (T_1,T_2,T_3) = e^{-(\gamma_e+\gamma_g+\Gamma_e/2)T_1-\Gamma_eT_2}\nonumber\\ \times e^{-(\gamma_e+\gamma_b+\Gamma_e/2+\Gamma_b/2)T_3}.
\end{gather}
This accounts not only for the dephasing and relaxation processes that affect the coherences during the waiting times $T_1$ and $T_3$ (as for the contributions related to ground state bleaching), but also for the relaxation taking place during the second waiting time $T_2$. }

\paragraph{\color{black} Non-rephasing contribution.}
The response function associated to the non-rephasing contribution corresponds to transitions 
$| 0 \rangle\langle 0 | \longrightarrow 
 | j \rangle\langle 0 | \longrightarrow 
 | j \rangle\langle k | \longrightarrow 
 | l \rangle\langle k | \longrightarrow 
 | k \rangle\langle k | $,
 where $|j\rangle$ and $|k\rangle$ are singly excited states, while $|l\rangle$ is doubly excited. In the case of model $A$, one has that $j=k=1$ and $l=3$. 
The expression of the response function reads:
\begin{gather}
    \mathcal{R}^{(3)}_6
    =i |\mu_{01}\mu_{23}|^2 \sum_{n_L,n_R=0}^\infty (-i|\eta|)^{2(n_L+n_R+1)}
    e^{h_M} \sum_{{\bf k}} \chi_M \nonumber\\  
    f_{M_L,{\bf q}_L,1} (\!-T_2\!-\!T_3)\, f_{M_R,{\bf q}_R,2} (T_1\!+\!T_2)\, 
    e^{-i (m_C \omega +\omega_3) T_3} .
\end{gather}
The $M$-dimensional vector ${\bf z}$ has the $l$-th component ${\rm z}_l=z_{j_l}$, where all the odd-numbered indices are $j_{2k+1}=1$, apart from $j_{M_L+1}=3$, and all the even-numbered are $j_{2k}=2$, apart from $j_0=j_{M+1}=0$. The overall order $M-3=2(n_L+n_R)+2$ in the non-adiabatic coupling results from $2n_R+1$ ($2n_L+1$) virtual transitions in the evolution of the ket (bra) during the first and second (second and third) waiting times. 

{\color{black} The effect of decoherence on the non-rephasing and rephasing contribution coincides. Therefore, also the above response function has to be multiplied by a factor $F_6(T_1,T_2,T_3)=F_3(T_1,T_2,T_3)$.}

\paragraph{\color{black} Verification against numerical results.}
In order to test these analytical results, we compare the third-order response function obtained for the rephasing contribution with that derived by numerical diagonalization of the Hamiltonian. As shown in Fig. \ref{fig:5}, the results of the perturbative approach (symbols) converge to the nonperturbative results (solid line) for increasing number of terms in the expansion. Terms of increasing order are clearly required for increasing values of $T_3$ and (not shown) of $T_1$, while the value of $T_2$ is irrelevant in this respect. 

\subsubsection{Double quantum coherence}

We  finally  consider  the  pathways  that  involve  coherences between the ground and a doubly excited state. These give rise to two kinds of contributions.

\paragraph{\color{black} First contribution.}
The response function associated to the first kind of contributions corresponds to transitions 
$| 0 \rangle\langle 0 | \longrightarrow 
 | j \rangle\langle 0 | \longrightarrow 
 | l \rangle\langle 0 | \longrightarrow 
 | l \rangle\langle k | \longrightarrow 
 | k \rangle\langle k | $,
 where $|j\rangle$ and $|k\rangle$ are singly excited states, while $|l\rangle$ is doubly excited. In the case of model $A$, one has that $j=k=1$ and $l=3$. The expression of the response function reads:
\begin{gather}
    \mathcal{R}^{(3)}_7
    =i |\mu_{01}\mu_{23}|^2 \sum_{n_L,n_R=0}^\infty (-i|\eta|)^{2(n_L+n_R+1)}
     e^{h_M} \sum_{{\bf k}} \chi_M  \nonumber\\ 
    f_{M_L,{\bf q}_L,1} (-T_3)\, f_{M_R,{\bf q}_R,2} (T_1)\, 
    e^{-i (m_C \omega +\omega_3) (T_2+T_3)} .
\end{gather}
The $M$-dimensional vector ${\bf z}$ has the $l$-th component ${\rm z}_l=z_{j_l}$, where all the odd-numbered indices are $j_{2k+1}=1$, apart from $j_{M_L+1}=3$, and all the even-numbered are $j_{2k}=2$, apart from $j_0=j_{M+1}=0$. The overall order $M-3=2(n_L+n_R)+2$ in the non-adiabatic coupling results from $2n_R+1$ ($2n_L+1$) virtual transitions in the evolution of the ket (bra) during the first (third) waiting time. 

{\color{black} Decoherence affects the above response function by inducing a decay of the single and double coherences that evolve during the three waiting times (Subsec. \ref{subsec:deco}). As a result, the above response function has to be multiplied by a factor $F_7$, whose expression reads:
\begin{gather}
    F_7 (T_1,T_2,T_3) = e^{-(\gamma_e+\gamma_g+\Gamma_e/2)T_1}\nonumber\\ \times e^{-(\gamma_b+\gamma_g+\Gamma_b/2)T_2-(\gamma_e+\gamma_b+\Gamma_e/2+\Gamma_b/2)T_3}.
\end{gather}}

\paragraph{\color{black} Second contribution.}
The response function associated to the second kind of contributions corresponds to transitions 
$| 0 \rangle\langle 0 | \longrightarrow 
 | j \rangle\langle 0 | \longrightarrow 
 | l \rangle\langle 0 | \longrightarrow 
 | k \rangle\langle 0 | \longrightarrow 
 | 0 \rangle\langle 0 | $,
where $|j\rangle$ and $|k\rangle$ are singly excited states, while $|l\rangle$ is doubly excited. In the case of model $A$, one has that $j=k=1$ and $l=3$. The expression of the response function reads:
\begin{gather}
    \mathcal{R}^{(3)}_8
    =-i |\mu_{01}\mu_{23}|^2 \sum_{n_L,n_R=0}^\infty (-i|\eta|)^{2(n_L+n_R+1)}
     e^{h_M} \nonumber\\ \sum_{{\bf k}} \chi_M  
    f_{M_L,{\bf q}_L,1} (T_3)\, f_{M_R,{\bf q}_R,2} (T_2)\, 
    e^{-i (m_C \omega +\omega_3) T_2}.
\end{gather}
The $M$-dimensional vector ${\bf z}$ has the $l$-th component ${\rm z}_l=z_{j_l}$, where all the odd-numbered indices are $j_{2k+1}=1$, apart from $j_{M_L+1}=3$, and all the even-numbered are $j_{2k}=2$, apart from $j_0=j_{M+1}=0$. The overall order $M-3=2(n_L+n_R)+2$ in the non-adiabatic coupling results from $2n_R+1$ ($2n_L+1$) virtual transitions in the evolution of the ket during the first (third) waiting time. 

{\color{black} The effect of decoherence on the second contribution that involves a double quantum coherence differs from that on the first contribution. In particular, the effect of dephasing and relaxation is accounted by a factor $F_8$, whose expression reads:
\begin{gather}
    F_8 (T_1,T_2,T_3) = e^{-(\gamma_e+\gamma_g+\Gamma_e/2)(T_1+T_3)} e^{-(\gamma_b+\gamma_g+\Gamma_b/2)T_2}.
\end{gather}}

{\color{black}

\subsection{Nonlinear response functions \\ in the frequency domain}

\begin{figure}[h]
\centering
\includegraphics[width=0.3\textwidth]{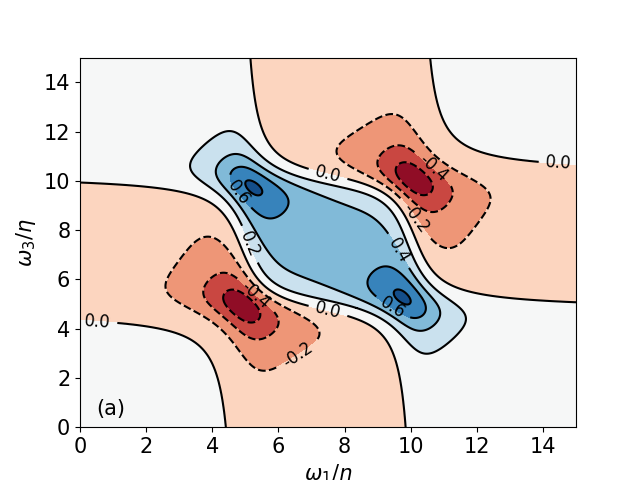}
\includegraphics[width=0.3\textwidth]{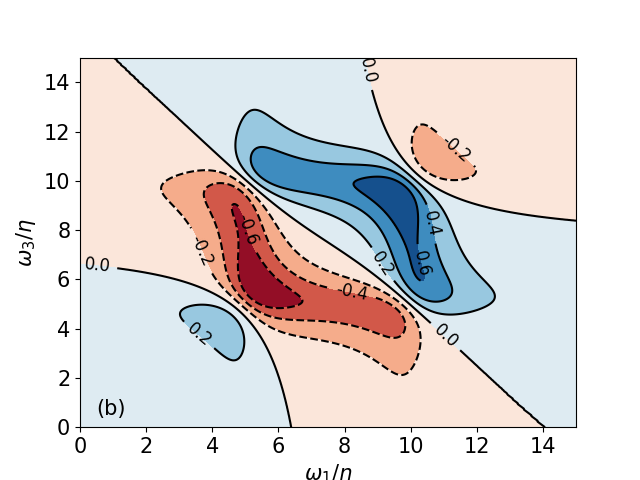}
\includegraphics[width=0.3\textwidth]{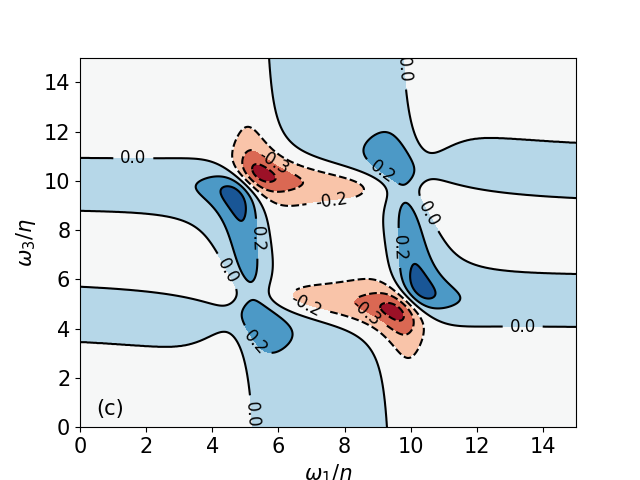}
\includegraphics[width=0.3\textwidth]{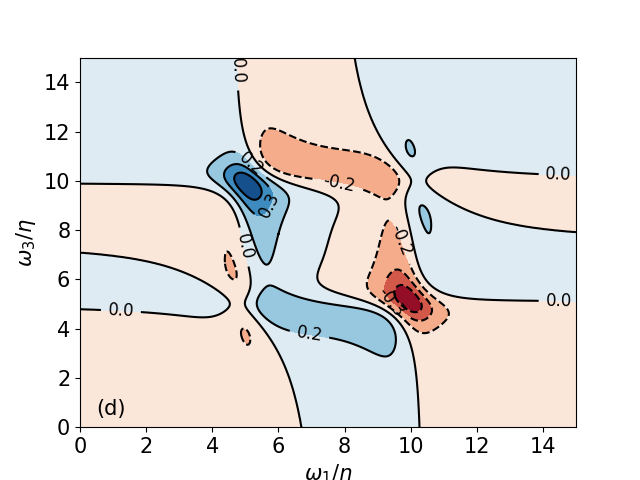}
\caption{\color{black} Real (a,c) and imaginary parts (b,d) of the second and fourth order contributions in the Dyson expansion of the third-order response function $\mathcal{R}_6^{(3)}/i|\mu_{01}\mu_{23}|^2$ (excited state absorption, non-rephasing contribution), Fourier transformed with respect to the waiting times $T_1$ and $T_3$, and for $T_2=0$. The model parameters are set to: $\bar\omega_1=5$,  $\bar\omega_2=10$,  $\bar\omega_3=15$, $\gamma_g=\gamma_e=\gamma_b=0.4$, and $\Gamma_e=\Gamma_b=0.4$, all in units of $\eta$, assumed to be real and positive.\label{fig:G}}
\end{figure}

The third-order response functions are given by the sum of terms corresponding to different orders $M_L+M_R-2$ in $\eta$. The same applies to the response functions in the frequency domain, $\mathcal{R}(\omega_1,T_2,\omega_3)$, obtained by performing the Fourier transform with respect to the times $T_1$ and $T_3$. 

In the following, we consider as a representative example the response function $\mathcal{R}_6^{(3)}$, related to excited state absorption, non-rephasing contribution, for $T_2=0$ and $z_1=z_2=0$ (Fig. \ref{fig:G}). We note in passing that, for $T_2=0$, this coincides with the response function related to the first contribution of the double quantum coherence, $\mathcal{R}_7^{(3)}$. 

The lowest nonzero contribution corresponds to $M_L=M_R=2$ (and thus to $|\eta|^2$). This physically corresponds to a single non-adiabatic transition $|1\rangle \longrightarrow |2\rangle$, taking place during the first two waiting times, and to a single non-adiabatic transition $\langle 1| \longrightarrow \langle 2|$, taking place during the last two waiting times. In the time domain, this term only includes  terms that oscillate at the diabatic states energies ($\Omega_{k,L}=\bar\omega_k$ and $\Omega_{k,R}=\bar\omega_{3-k}$, with $k=1,2$), with constant prefactors $A_1$ and $A_2$.
The resulting contribution [panels (a) and (b), real and imaginary parts, respectively] is characterized by the presence of two identical diagonal peaks at the diabatic states energies, and by off-diagonal peaks with opposite sign. 

The following nonzero contribution corresponds to $M_L=M_R=8$ (and thus to $|\eta|^6$). This physically corresponds to the triple transition $|1\rangle \longrightarrow |2\rangle \longrightarrow |1\rangle \longrightarrow |2\rangle$, taking place during $T_1+T_2$, and to the transition $\langle 1| \longrightarrow \langle 2|\longrightarrow \langle 1|\longrightarrow \langle 2|$, taking place during $T_2+T_3$. 
In the time domain, this term still includes  terms that oscillate at the diabatic states energies ($\bar\omega_1$ and $\bar\omega_2$), but with prefactors that are linear in the relevant waiting times ($T_1$ and $T_3$). The resulting contribution [panels (c) and (d), real and imaginary parts, respectively] is characterized by the presence of more complex features in the diagonal and off-diagonal positions, with hybrid absorptive and dispersive character (see Subsec. \ref{subsec:Pitsd}).  
}

\section{Derivations \label{sec:derivations}}

In the following, we provide the formal derivation of the results reported in Sec. \ref{sec:main results}.

\subsection{Time-evolution operator}

The starting point is the introduction of an interaction picture, based on the separation of the adiabatic and non-adiabatic components of the Hamiltonian: $H=(H_g+H_{0,e}+H_b)+V\equiv H_0+V$. The terms corresponding to the ground ($H_g$) and doubly-occupied states ($H_b$) are by assumption adiabatic, while the projection of the Hamiltonian onto the subspace $\mathcal{S}_e = \{|1\rangle,|2\rangle \}$ includes both an adiabatic ($H_{0,e}$) and a non-adiabatic ($V$) term. Hereafter, the focus is on the free dynamics that takes place within the subspace $\mathcal{S}_e$, which can undergo optical transitions from and to the ground- and doubly-occupied states.

In the interaction picture, the time-dependent state is given by: $|\psi_I(t)\rangle = e^{iH_0t} e^{-iHt} |\psi (0)\rangle \equiv U_I |\psi (0)\rangle$, where the time evolution operator reads \cite{Mahan}:
\begin{align}\label{eq:02}
    U_I(t) \! = \! 1 \! +\sum_{n=1}^\infty (-i)^n\!  \int^t_0 dt_1 \dots\int^{t_{n-1}}_0 dt_n V_I(t_1)\dots V_I(t_n) .
\end{align}
From this, one can obtain the time evolution operator in the Schr\"odinger picture: $U_S=e^{-iHt}=e^{-iH_0t} U_I$.

The non-adiabatic operator corresponds to $V_I (t) = e^{iH_0t} V e^{-iH_0t} = e^{iH_{0,e}t} V e^{-iH_{0,e}t}$. The exponential operators can be expressed in terms of the displacement operators $\mathcal{D} (\alpha) = e^{\alpha a^\dagger - \alpha^* a}$ as follows:
\begin{align}
    e^{\pm i H_{0,e} t} = \sum_{\sigma=1}^2 | \sigma \rangle\langle \sigma | \, e^{\pm i\bar\omega_\sigma t}\, \mathcal{D} (-z_\sigma)\, e^{\pm i \omega a^\dagger a t}\, \mathcal{D} (z_\sigma) ,
\end{align}
being $z_\sigma$ the displacement corresponding to the electronic state $\sigma$. 
From this it follows that the non-adiabatic component of the Hamiltonian is given by:
\begin{align}
    V_I (t) & = \eta |1\rangle\langle 2| \, e^{i\bar\omega_{12} t}\, \mathcal{D} (-z_1)\,  e^{i \omega a^\dagger a t}\,  \nonumber\\ &  \mathcal{D} (z_{12})\,  e^{- i \omega a^\dagger a t}\,  \mathcal{D} (z_2) + {\rm H.c.},
\end{align}
where $\bar\omega_{12}\equiv\bar\omega_1-\bar\omega_2=-\bar\omega_{21}$.

The products of an odd number of operators $\hat{V}$ that appear in Eq. (\ref{eq:02}) can thus be written as
\begin{align}\label{eq:03}
    V_I & (t_1) \dots V_I (t_{2n+1})  = \eta |\eta|^{2n} |1\rangle\langle 2| e^{i\bar\omega_{21} \sum_{k=1}^{2n+1} (-1)^{k} t_k} \nonumber\\ &\mathcal{D} (-z_2) \left\{\prod_{l=1}^{2n+2} \mathcal{D} [(-1)^{l} z_{12}] e^{-i \omega a^\dagger a \tau_l} \right\}  \mathcal{D} (z_2) + {\rm H. c.} ,
\end{align}
where $t_0=t_{2n+2}=0$. They physically correspond to contributions where the system undergoes $2n+1$ transitions between the states $|1\rangle$ and $|2\rangle$, at the times $t_{2n+1}<t_{2n}<\dots<t_1$, {\color{black} separated by time intervals of duration $\tau_k=t_{k-1}-t_k$}.

The products of an even number of non-adiabatic operators are diagonal in the basis of the adiabatic states and read:
\begin{align}\label{eq:04}
    V_I & (t_1) \dots V_I (t_{2n}) = |\eta|^{2n} |1\rangle\langle 1| e^{i\bar\omega_{21} \sum_{k=1}^{2n} (-1)^{k} t_k} \nonumber\\ & \mathcal{D} (-z_2) \left\{\prod_{l=1}^{2n+1} \mathcal{D} [(-1)^{l} z_{12}] e^{-i \omega a^\dagger a \tau_l}\right\} \mathcal{D} (z_1) \nonumber\\ & + 
    |\eta|^{2n} |2\rangle\langle 2| e^{i\bar\omega_{12} \sum_{k=1}^{2n} (-1)^{k} t_k} \nonumber\\ & \mathcal{D} (-z_1) \left\{\prod_{l=1}^{2n+1} \mathcal{D} [(-1)^{l} z_{21}] e^{-i \omega a^\dagger a \tau_l}\right\} \mathcal{D} (z_2)  ,
 \end{align}
 where $t_0=t_{2n+1}=0$. They physically correspond to contributions where the system undergoes $2n$ transitions between the states $|1\rangle$ and $|2\rangle$, at the times $t_{2n}<t_{2n-1}<\dots<t_1$, {\color{black} separated by time intervals of duration $\tau_k=t_{k-1}-t_k$}.
 
\subsection{Propagators at defined interaction times}

From the above equations it follows that the matrix element between the electronic states $|\sigma\rangle$ and $|\sigma'\rangle$ (where $\sigma,\sigma'=1,2$) of the products $e^{-i H_{0,e}t} V_I (t_1)\dots V(t_{M-1})$ can always be written as alternating sequences of displacement operators and free-oscillator time-evolution operators. The expectation value of such operators in the vacuum state $|0\rangle$ of the vibrational mode, to which we refer in the following as {\it adiabatic response function}, has a well defined analytical expression, which reads \cite{Quintela2022a}:
\begin{align}\label{eq:arf}
   R^{(v,M)}_{j_1\,j_M} & (\tau_1,\dots,\tau_M) = \exp[h_M({\bf z})] \times\nonumber\\
   & \exp \left(-\sum_{k=1}^M \sum_{l=1}^{M-k+1} z_{j_{l-1},j_l} z_{j_{l+k-1},j_{l+k}} \prod_{p=l}^{l+k-1} v_p\right).
\end{align}
The function $h_M$ of the displacements is given in Eq. (\ref{eq:hM}).
We stress that $R_{j_1\,j_M}^{(v,M)}$ formally coincides with the vibrational response function for the displaced harmonic oscillator model, but has here a different physical interpretation. In particular, the transition between electronic states were induced there by the interaction of the system with the electric field, and here by the nonadiabatic term $V$. In order to stress such difference, the response function for the non-adiabatic model that is considered in the present paper is denoted with the symbol $\mathcal{R}$.

The adiabatic response function $R_{j_1\,j_M}^{(v,M)}$ can be associated to a time evolution of the vibrational state induced by an Hamiltonian that is piece wise constant, and undergoes abrupt transitions as the system undergoes transitions between the electronic state $|1\rangle$ or $|2\rangle$. In particular, the Hamiltonian is constant during each of the $M$ time intervals, $\tau_k=t_{k-1}-t_k$ ($k=1,\dots,M$), delimited by two consecutive transitions. 
At each time interval one can associate a function $v_k=e^{-i\omega\tau_k}$, which appears in the expression of $R_{j_1\,j_M}^{(v,M)}$, and an index $j_k=1,2$, which specifies the electronic state and thus the Hamiltonian $H_{0,j_k}$ that induces the time evolution. The index $j_k$ also specifies the relevant displacement $z_{j_k}$, whose differences $z_{j_{k-1},j_k}\equiv z_{j_{k-1}} - z_{j_{k}}$ appear in Eq. (\ref{eq:arf}). 

\begin{figure}[h]
\centering
\includegraphics[width=0.4\textwidth]{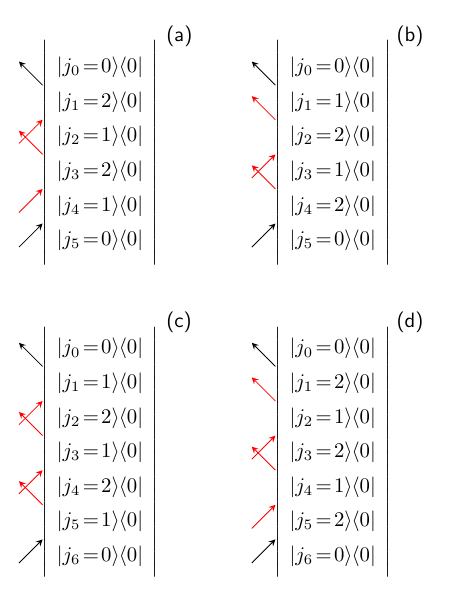}
\caption{Representation in terms of the double-sided Feynman diagrams of the adiabatic response functions $R^{(v,4)}_{21}$ (a), $R^{(v,4)}_{12}$ (b), $R^{(v,5)}_{11}$ (c), $R^{(v,5)}_{22}$ (d). The first and last (black) arrows correspond to transitions from and to the electronic ground state $|0\rangle$, induced by the electric field. The intermediate (red) arrows represent transitions between the excited states $|1\rangle$ and $|2\rangle$, induced by the non-adiabatic term $V$ in the Hamiltonian. \label{fig:A}}
\end{figure}

In the case of products of odd-order terms, the expectation value that enters the expression of the matrix element $\langle 1;0 | U_S | 2;0 \rangle $ reads:
\begin{gather}
    \langle 1;0 | e^{-iH_{0,1} t} V_I (t_1)\dots V_I(t_{M-1}) |2; 0\rangle = e^{-i\bar\omega_1 t} \nonumber\\ 
    \label{eq:100} \eta |\eta|^{2n} e^{i\bar\omega_{21} \sum_{k=1}^{2n+1} (-1)^{k} t_k} R^{(v,M)}_{12} (\tau_1,\dots,\tau_{M})
\end{gather}
where $M=2n+2$, $j_{2k}=2$, $j_{2k+1}=1$, apart from $j_0=j_{M+1}=0$.
The corresponding Feynman diagrams are characterized by $M+1$ arrows, all on the left side [Fig. \ref{fig:A}(b)], with the state before the first interaction and after the last one both coinciding with $|0\rangle$, and in between an alternation of states $|1\rangle$ and $|2\rangle$.
The time-independent term in the exponent of the adiabatic response function is given by $h_M({\bf z}) = -\frac{1}{2} M z_{12}^2 - z_1 z_2 $. 
The expression of the propagator $\langle 2,0 | e^{-iH_{0,2} t} V_I (t_1)\dots V_I(t_{M-1}) |1, 0\rangle$ [see Fig. \ref{fig:A}(a)] can be obtained from the above expression by swapping the indices 1 and 2 that define the electronic states and by replacing $\eta$ with its complex conjugate.

In the case of even-order terms, the expectation value that enters the expression of the matrix element $\langle \sigma ;0 | U_S | \sigma ;0 \rangle $ reads:
\begin{gather}
    \langle \sigma; 0 | e^{-iH_{0,\sigma} t} V_I (t_1)\dots V_I (t_{M-1}) |\sigma; 0\rangle = e^{-i\bar\omega_\sigma t} \nonumber\\ \label{eq:200} |\eta|^{2n} e^{i\bar\omega_{12} \sum_{k=1}^{2n} (-1)^{k+\sigma} t_k}  R^{(v,M)}_{\sigma\sigma} (\tau_1,\dots,\tau_{M})
\end{gather}
where $\sigma=1,2$, $M=2n+1$, $j_{2k}=3-\sigma$, $j_{2k+1}=\sigma$, apart from $j_0=j_{M+1}=0$. The corresponding Feynman diagrams are characterized by $M+1$ arrows, all on the left side [Fig. \ref{fig:A}(c,d)], with the state before the first interaction and after the last one both coinciding with $|0\rangle$, in between an alternation of $|1\rangle$ and $|2\rangle$. 
The time-independent term in the exponent of the adiabatic response function is given by $h_M({\bf z}) = -\frac{1}{2} (M-1) z_{12}^2 - z_\sigma^2 $. 

\subsection{Taylor expansion of the propagator \label{subsec:te}}

In order to compute the integrals with respect to the interaction times, we expand the response functions $R_{\sigma\sigma'}^{(v,M)}$ in Taylor series with respect to all the exponentials that appear in the exponent. In particular, the response function of order $M$ is given by the sum of $M(M+1)/2$ terms: the first $M$ terms 
$v_1,v_{2},\dots,v_M$ correspond to the individual time intervals $\tau_1,\tau_{2},\dots,\tau_{M}$, the following $M-1$ terms $v_1 v_{2}, v_{2} v_{3},\dots,v_{M-1} v_M$ correspond to the double time intervals $\tau_1 + \tau_{2},\tau_{2}+\tau_{3},\dots,\tau_{M-1}+\tau_{1}$; and so on until the last term $v_1 v_2 \dots v_{M-1} v_M$, which corresponds to the $M$-tuple time interval $t=\sum_{k=1}^M\tau_M$. The Taylor expansion thus gives:
\begin{gather}
    R_{\sigma\sigma'}^{(v,M)}=e^{h_M({\bf z})}\sum_{\bf k} \frac{(z_\sigma z_{\sigma\bar\sigma})^{k_1}\dots (z_\sigma z_{\sigma'})^{k_{M(M+1)/2}}}{k_1!\dots k_{M(M+1)/2}!}\,
    \nonumber\\ e^{-i\omega k_1 t_{M-1}} \dots e^{-i\omega k_{M+1} t_{M-2}} \dots e^{-i\omega k_{M(M+1)/2} t} \nonumber\\  \label{eq:expansion} 
    = e^{h_M} \sum_{\bf k} \chi_M \prod_{p=0}^{M-1} e^{-i\omega t_{p} m_{p}} 
    = e^{h_M} \sum_{\bf k} \chi_M \prod_{p=1}^{M} e^{-i\omega \tau_{p} q_{p}} ,
\end{gather}
where ${\bf z}=(z_1,\dots,z_M)$ and ${\bf k}=[k_1,\dots,k_{M(M+1)/2}]$, with the components $k_i$ that vary from $0$ to $\infty$.

In the last equation above, the $M(M+1)/2$ oscillating terms are reduced either to the $M$ terms that depend on one of the times $t_p$, or to the $M$ that depend on the time intervals $\tau_p$. In the former case, the exponent of each factor in the last line above, $m_{p}=l_{M-p}-l_{M-p+1}$ ($p=0,\dots,M-1$), depends on ${\bf k}$ through the expression 
\begin{align}\label{eq:lp}
l_{M-p}\! =\! k_{M-p}\! + \sum_{q=2}^{M} \sum_{s=\max(0,q-p-1)}^{\min(q-1,M-p-1)} \!\!\!\!\!\!\!\!\! k_{q M-p-(q-1)(q-2)/2-s} .
\end{align}
The coefficients $q_j$ can be expressed as a function of the $m_i$, being $q_j=\sum_{i=0}^j m_i$.
The function $\chi_M$ depends both on ${\bf z}$ and on ${\bf k}$, through the expression reported in Eq. (\ref{eq:chiM}). As a result, the propagator corresponding to defined interaction times is expressed as sum of terms, each one given by a product of exponential functions of the times.

\subsection{Integration over the interaction times \label{subsec:iit}}

In order to derive the matrix elements of the time-evolution operator $U_S$, one finally needs to integrate the above quantities, multiplied by the additional oscillating terms [see Eqs. (\ref{eq:03},\ref{eq:04})], with respect to the $M-1$ times $t_p$. The multiple integral gives rise to the following expression:
{\color{black}
\begin{gather}
    f_{M,{\bf q},\sigma}(t) = e^{-i(\omega m_0+\bar\omega_\sigma)t} \prod_{p=1}^{M-1} \int_0^{t_{p-1}} dt_{p}\, e^{i\omega_{pp} t_{p}} \nonumber\\ 
    = \sum_{j=1}^{M} A_{j}(t)\, e^{-i(\bar\omega_\sigma-\omega_{0,j-1})\, t}
    \equiv  \sum_{j=1}^{M} A_{j}(t)\, e^{-i\Omega_{j} t} \label{eq:05} 
\end{gather}
where ${\bf q}\equiv (q_1,\dots,q_{M})$. Besides, $A_{j}(t) = a_{j} t^{r_j}$, being $r$ the number of zero frequencies amongst the $\omega_{k,j-1}$, for $k=1,\dots,j-1$. 

Besides the $\Omega_j$, which appear in the final expression above, it is thus necessary to introduce the frequencies $\omega_{kj}$, which take the value
\begin{align}\label{eq:mfr}
    \omega_{kj} = - \omega\sum_{i=k}^j m_i - \frac{1}{2}[(-1)^{k+\sigma}+(-1)^{j+\sigma}]\,\bar\omega_{21} 
\end{align}
for $k\le j$ and $\omega_{kj}=0$ for $k>j$. From the above expression of the $\omega_{kj}$ it follows that $r$ cannot be larger than $M/2-1$, for even values of $M$, and of $(M-1)/2$, for odd values. The frequencies $\omega_{kj}$ and $\Omega_j$ can be expressed as a function of one another, through the relations:
\begin{align}
    \Omega_j = \bar\omega_\sigma-\omega_{0,j-1},\ \omega_{kj}=\Omega_{k}-\Omega_{j+1}
\end{align}}

If none of the frequencies $\omega_{kj}$ vanishes, then one can define a set of constants $A_{M-1-k,j}(t)$, with $k = 1, \dots ,M – 1$ and $j = 1, \dots , k + 1)$. By sequentially performing the integrals in Eq. (\ref{eq:05}), one can show that the following recurrence relations apply, starting from 
$A_{M-1,1}=1$:
\begin{gather}
    A_{M-1-k,j} = \frac{A_{M-k,j-1}}{i\omega_{M-k,M-2-k+j}}\\
    A_{M-1-k,1}=-\sum_{j=2}^{k+1} A_{M-1-k,j} .
\end{gather}
Combining together the above equations, one can eventually express all the coefficients that enter the expression of the functions $f_{M,{\bf q},\sigma}(t)$ in terms of the frequencies $\omega_{ij}$:
\begin{gather}\label{eq:frequencies}
    A_{M-m} =  \frac{(-1)^m\,i^{1-M}}{\prod_{k=1}^{M-m-1} \omega_{k,M-m-1}\prod_{l=M-m}^{M-1} \omega_{M-m,l}} ,
\end{gather}
being $A_{M-m} \equiv A_{0,M-m}$.

In the presence of zero frequencies, the above recursive relations have to be modified. One can derive Eq. (\ref{eq:05}) by introducing functions $A_{ij}(t)=\sum_{k=0}^r a_{ijk} t^k$. If $\omega_{M-k,M-k+j-2}\neq 0$, then
\begin{gather}
    a_{M-k-1,j,r} = \sum_{s=r}^{b} \frac{s!}{r!} \frac{(-1)^{s-r}\, a_{M-k,j-1,s}}{(i\omega_{M-k,M-k+j-2})^{s-r+1}},
\end{gather}
where $b$ is the order of the polynomial $A_{M-k,j-1}$,
and the constant term in the polynomial is given by
\begin{gather}
    a_{M-k-1,1,0}=-\sum_{j=2}^{k+1} a_{M-k-1,j,0} .
\end{gather}
If instead $\omega_{M-k,M-k+j-2} = 0$, then
\begin{gather}
    a_{M-1-k,j,r} = \frac{1}{r}\, a_{M-k,j-1,r-1}
\end{gather}
and $a_{M-k-1,j,0}=0$.

\begin{figure}[h]
\centering
\includegraphics[width=0.4\textwidth]{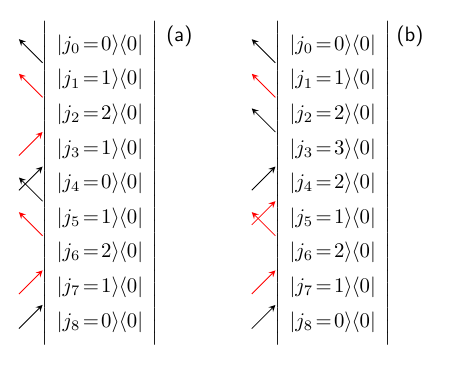}
\caption{Representation in terms of the double-sided Feynman diagrams of the third-order response functions of the type $\mathcal{X}_1$ (a) and $\mathcal{X}_2$ (b). The black arrows correspond to transitions from and to the electronic ground ($|0\rangle$) or doubly-excited ($|2\rangle$) states, induced by the field. The red arrows denote transitions between the excited states $|1\rangle$ and $|2\rangle$, induced by the non-adiabatic term $V$ in the Hamiltonian. \label{fig:B}}
\end{figure}

{\color{black}
\subsection{Decoherence\label{subsec:deco}}

The effect of decoherence can be included in the present approach at a phenomenological level. In particular, such inclusion leads to simple time-dependent prefactors for the derived response functions under the condition that the environment couples symmetrically to the subspace $\mathcal{S}_{e}=\{|1\rangle,|2\rangle\}$ where the non-adiabatic term is defined. This implies that pure dephasing between $|1\rangle$ and $|2\rangle$ is not included, and that these two states are assumed to relax at an equal rate. 

In the presence of decoherence, the free evolution of the system between two consecutive transitions induced by the electric field can no longer be simulated by the Schr\"odinger equation. We thus refer to a master equation in the Lindblad form \cite{Breuer},
\begin{gather}
    \frac{d}{dt} \rho = i [\rho,H] + \sum_{i=1}^{N_L} \left[L_i \rho L_i^\dagger - \frac{1}{2} (L_i^\dagger L_i \rho + \rho L_i^\dagger L_i)\right],
\end{gather}
with $N_L=6$ Lindblad operators $L_i$. Three of these, namely 
\begin{gather}\label{lo1}
L_1=\sqrt{\Gamma_e}\, |0\rangle\langle 1|,\ L_2=\sqrt{\Gamma_e}\, |0\rangle\langle 2|,\ L_3=\sqrt{\Gamma_b}\, |0\rangle\langle 3|,
\end{gather}
account for relaxation, respectively from the states $|1\rangle$, $|2\rangle$, and $|3\rangle$. The other three operators read:
\begin{gather}
L_4=\sqrt{2\gamma_g}\, |0\rangle\langle 0|,\ L_6=\sqrt{2\gamma_b}\, |3\rangle\langle 3|, \nonumber\\ \label{lo2}
L_5=\sqrt{2\gamma_e}\, (|1\rangle\langle 1|+|2\rangle\langle 2|),
\end{gather}
and account respectively for the decay of coherences between the subspaces $\mathcal{S}_g=\{|0\rangle\}$, $\mathcal{S}_b=\{|3\rangle\}$, and $\mathcal{S}_e$, and any other subspace. It should be intended that each of the above operators $L_i$ is multiplied by an identity operator that applies to the vibrational degrees of freedom, and thus has no direct effect of the state of the harmonic oscillator(s). 

The coherences between states belonging to different subspaces decay at a rate which is given by the sum of the respective dephasing rates and of the average relaxation rate. For example, 
$\dot\rho_{13}=-\rho_{13}[\gamma_e+\gamma_b+\frac{1}{2}(\Gamma_e+\Gamma_b)+i\bar\omega_{13}]$, with $\bar\omega_{ij}\equiv\bar\omega_i-\bar\omega_j$.
Coherences between the states $|1\rangle$ and $|2\rangle$, instead, undergo an exponential decay only in virtue of the relaxation from the subspace $\mathcal{S}_e$: $\dot\rho_{12}=-\rho_{12}(\Gamma_e+i\bar\omega_{12})$. 

The same exponential decay affects the populations $\rho_{11}$ and $\rho_{22}$. As a result, the superoperators associated with all the Lindblad operators $L_i$ commutes with the one related to the Hamiltonian, and the effect of decoherence on the time evolution of any $\rho_{ij}$ can be reduced to a multiplicative exponential decay, with suitable decay rate. This set of Lindblad operators doesn't account for a pure dephasing term within the subspace $\mathcal{S}_3$. Its inclusion would require a generalization of the derivations presented in Sec. \ref{sec:derivations}, which is beyond the scope of the present article.  

In view of the above results, the effect of the Lindblad operators reported in Eqs. (\ref{lo1}-\ref{lo2}) can be effectively incorporated in the expression of the propagators and of the response function, through the inclusion of prefactors that decay exponentially with the waiting times. We note for completeness, that this approach accounts for the effects of the population loss in the initial state of the relaxation process, but not for those of the population gain in the final state.}

\subsection{Multi-mode case}

The above results can be generalized to the case of multiple ($G >1$) vibrational modes. 
The procedure is the one that has been followed in the single-mode case: calculation of the operators $V_I(t)$ and of their products; identification of their expectation values in the ground state of the vibational modes with the multimode adiabatic response functions; integration with respect to the interaction times. In the case of products of odd-order terms, such expectation value reads:
\begin{gather}
    \langle 1;{\bf 0} | e^{-iH_{0,1} t} V_I (t_1)\dots V_I(t_{2n+1}) |2; {\bf 0}\rangle = e^{-i\bar\omega_1 t} \nonumber\\ 
     \eta |\eta|^{2n} e^{i\bar\omega_{21} \sum_{k=1}^{2n+1} (-1)^{k} t_k} \prod_{\zeta =1}^G R^{(v_\zeta,M)}_{12} (\tau_1,\dots,\tau_M)
\end{gather}
where, $|{\bf 0}\rangle\equiv |0,\dots,0\rangle$ is the multimode ground state. As in the case $G=1$, the following relations hold: $M=2n+2$, $j_{2k}=2$, $j_{2k+1}=1$, apart from $j_0=j_{M+1}=0$. Physically, this term still refers to the occurrence of $2n+1$ hopping processes between the excited states, at the times $t_{2n+1} < t_{2n} < \dots < t_1 $, which eventually lead to a transition from $|2\rangle$ to $|1\rangle$.

In the case of even-order terms, the expectation value of the vibrational ground state reads:
\begin{gather}
    \langle \sigma; {\bf 0} | e^{-iH_{0,\sigma} t} V_I (t_1)\dots V_I (t_{2n}) |\sigma; {\bf 0} \rangle = e^{-i\bar\omega_\sigma t} \nonumber\\ |\eta|^{2n} e^{i\bar\omega_{12} \sum_{k=1}^{2n} (-1)^{k+\sigma} t_k} \prod_{\zeta =1}^G R^{(v_\zeta,M)}_{\sigma\sigma} (\tau_1,\dots,\tau_M)
\end{gather}
where, as in the case $G=1$, $\sigma=1,2$, $M=2n+1$, $j_{2k}=3-\sigma$, $j_{2k+1}=\sigma$, apart from $j_0=j_{M+1}=0$. Physically, this term refers to the occurrence of $2n$ hopping processes between the states $|2\rangle$ to $|1\rangle$, at the times $t_{2n} < t_{2n} < \dots < t_1 $, which eventually bring the system back to its initial state.

We are now in the condition of writing the final expression of the propagators. In particular, the off-diagonal one in the basis $\{|1\rangle,|2\rangle\}$ reads:
\begin{gather}
    \langle 1; {\bf 0} | U_S | 2; {\bf 0} \rangle = \sum_{n=0}^\infty (-i)^{2n+1}\eta |\eta|^{2n} \nonumber\\ \left[ \prod_{\zeta=1}^G e^{h_{M}({\bf z_\zeta})} \sum_{{\bf k}_\zeta} \chi_{M} ({\bf z}_\zeta,{\bf k}_\zeta) \right]\,  
    f_{M,{\bf Q},1} (t),
\end{gather}
where $M=2n+2$, $j_{2k}=2$, $j_{2k+1}=1$, apart from $j_0=j_{M+1}=0$. Besides, ${\bf Q} \equiv ({\bf q}_1,\dots,{\bf q}_G) $, where the relation between the vector ${\bf q}_\zeta$ and ${\bf k}_\zeta$ is given by Eq. (6).

The diagonal part of the propagator is given by the following expression:
\begin{gather}
    \langle \sigma; {\bf 0} | U_S | \sigma; {\bf 0} \rangle = \prod_{\zeta=1}^G u_{0,\zeta} + \sum_{n=1}^\infty (-i)^{2n} |\eta|^{2n} \nonumber\\ \left[ \prod_{\zeta=1}^G e^{h_{M}({\bf z_\zeta})} \sum_{{\bf k}_\zeta} \chi_{M} ({\bf z}_\zeta,{\bf k}_\zeta) \right]\, f_{M,{\bf q},\sigma} (t)
\end{gather}
where $M=2n+1$, $j_{2k}=3-\sigma$, $j_{2k+1}=\sigma$, apart from $j_0=j_{M+1}=0$, and $u_{0,\zeta}=\exp[z_{1,\zeta}^2(e^{-i\omega_\zeta t}-1)] e^{-i\bar\omega_\sigma t}$. As in the even-$M$ case, the time-dependent polynomials are obtained from the single-mode expressions by replacing $\omega q_p$ with $\sum_{\zeta=1}^G \omega_\zeta q_{p,\zeta}$.

A simple and yet relevant case is one where the two excited states correspond to electronic excitations localized in the first or second component of a dimer: $|1\rangle = |e,g\rangle$ and $|2\rangle=|g,e\rangle$ (model $B$, Fig. \ref{fig:2}). The model includes two vibrational modes ($G=2$), each one localized in one of the monomers. The oscillator displacement vanishes when the corresponding monomer is in the ground state ($z_{\zeta=1,2}=z_{\zeta=2,1}=0$).
If the two units are identical, then the two vibrational frequencies and the displacements ($z_e\equiv z_{\zeta=1,1}=z_{\zeta=2,2}\neq 0$) coincide, and $\bar\omega_{12}=0$. In this case, the two-mode adiabatic response function can be written as a single-mode one, by replacing $h_M({\bf z})$ and $\chi_M({\bf z},{\bf k})$ respectively with $h_M'({\bf z})$ and $\chi_M'({\bf z},{\bf k})$. In particular, one can show that $h_{M}'({\bf z}) = - Mz_e^2 $ for all values of $\sigma,\sigma'=1,2$. 
As to the functions $\chi_M' ({\bf z},{\bf k})$ [Eqs. (8) and (13)], their nominators are given by products of terms $X_i^{k_i}$. The terms corresponding to $i=(p-1)M-(p-1)(p-2)/2+1$ and $i=pM-p(p-1)/2$ (with $p = 1, \dots ,M - 1$) is $X_i=(-1)^{p+1}z^2_e$, while the term corresponding to $i=M(M+1)/2$ is $z^2_e$ for $\sigma=\sigma'$. and 0 otherwise; in all the other cases, $X_i= 2(-1)^{p+1} z^2_e$.

\subsection{Multitime propagators \\ and nonlinear response functions}

The present approach can also be applied to multitime propagators, such as the ones that enter the expressions of nonlinear response functions. We focus hereafter on the three-time propagators, which typically represents the most relevant one in multidimensional coherent spectroscopy. 

For the sake of simplicity, we consider the case where optical transitions are only allowed between the ground state $|0\rangle$ and the excited state $|1\rangle$, and between $|2\rangle$ and the doubly-excited state $|3\rangle$ (model $A$, Fig. \ref{fig:1}). The relevant and inequivalent propagators can thus be reduced to two. In the first one, the left and right propagators only involve the state $|1\rangle$, while the central one involves the ground state: 
\begin{gather}
    \mathcal{X}_1 = \langle 1;0| e^{-i H T_L} | 1\rangle \langle 0 | e^{-i H T_C} | 0 \rangle \langle 1 | e^{-i H T_R} | 1;  0\rangle .
\end{gather}

In order to derive the above quantities, one can proceed along the same lines as for the single-time propagators. In a first step, the time evolution operators associated to the non-adiabatic Hamiltonian $H_e$ are expanded in powers of $V_I$. As a result, one has, for given values of the intermediate times $t_{L,2n_L}<t_{L,2n_L-1}<\dots,t_{L,1}$ and $t_{R,2n_R}<t_{R,2n_R-1}<\dots <t_{R,1}$ an operator given by an alternating sequence of displacement operators and free oscillator time evolution operators:
\begin{gather}
    \langle 1 ; 0 | e^{-iH_{0,1} T_L} V_I (t_{L,1})\dots V_I(t_{L,2n_L}) |1\rangle \langle 0| e^{-i H_{0,0} T_C} |0\rangle \nonumber\\ 
    \langle 1 | e^{-iH_{0,1} T_R} V_I (t_{R,1})\dots V_I(t_{R,2n_R}) |1; 0\rangle  \nonumber\\ 
    =  e^{-i\bar\omega_1(T_L+T_R)} |\eta|^{2(n_L+n_R)} e^{i\bar\omega_{21}\sum_{\xi=L,R} \sum_{k=1}^{2n_\xi} (-1)^k t_{\xi,k} }  \nonumber\\ R^{(v,M)}_{11} (\tau_{L,1},\dots,\tau_{L,M_L},T_C,\tau_{R,1},\dots,\tau_{R,M_R}),
\end{gather}
where $M_L=2n_L+1$ and $M_R=2n_R+1$.
This can be formally identified with an adiabatic response function of order $ M = 2(n_L + n_R) + 3$, where and all the odd-numbered indices are $j_{2k+1}=1$ and all the even-numbered are $j_{2k}=2$, apart from $j_0=j_{M_L+1}=j_{M+1}=0$.

In a second step, the adiabatic response function is expanded in powers of the exponentials that appear in the exponent. Finally, the multiple integration is performed independently with respect to the interaction times $t_{L,i}$ and $t_{R,j}$. As a result, one obtains
\begin{gather}
    \mathcal{X}_1 
    =\sum_{n_L,n_R=0}^\infty (-i|\eta|)^{2(n_L+n_R)}  e^{h_M} \nonumber\\ \sum_{{\bf k}} \chi_M\, 
    f_{M_L,{\bf q}_L,1} (T_L)\, f_{M_R,{\bf q}_R,1} (T_R)\, 
    e^{-i m_C \omega T_C}.
\end{gather}

In the second case, the system undergoes a transition from $|1\rangle$ to $|2\rangle$ during the time $T_R$. The state occupied during $T_C$ necessarily coincides with $|3\rangle$, being this the only electronic state that is optically coupled to $|2\rangle$: 
\begin{gather}
    \mathcal{X}_2 = \langle 1;0 | e^{-i H T_L} | 2\rangle \langle 3 | e^{-i H T_C} | 3 \rangle \langle 2 | e^{-i H T_R} | 1; 0\rangle .
\end{gather}
The expansion with respect to the non-adiabatic term $V_I$, where now only odd powers contribute, leads to:
\begin{gather}
    \langle 1;0 | e^{-iH_{0,1} T_L} V_I (t_{L,1})\dots V_I(t_{L,2n_L+1}) |2\rangle \langle 3 | e^{-i H_{0,3} T_C} | 3 \rangle \nonumber\\ 
    \langle 2 | e^{-iH_{0,2} T_R} V_I (t_{R,1})\dots V_I(t_{R,2n_R+1}) |1; 0\rangle =  e^{-i\bar\omega_1 T_L} \nonumber\\ e^{-i (\omega_3 T_C + \bar\omega_2 T_R)}
    |\eta|^{2(n_L+n_R+1)} e^{i\bar\omega_{21}\sum_{\xi=L,R}\sum_{k=1}^{2n_\xi+1} (-1)^k t_{\xi,k}} \nonumber\\ R^{(v,M)}_{11} (\tau_{L,1},\dots,\tau_{L,M_L},T_C,\tau_{R,1},\dots,\tau_{R,M_R}) ,
\end{gather}
where $M_L=2(n_L+1)$ and $M_R=2(n_R+1)$.
In the adiabatic response function or order $M=2(n_L+n_R)+5$, all the odd-numbered indices are $j_{2k+1}=1$, apart from $j_{M_L}=3$, and all the even-numbered are $j_{2k}=2$.  

After performing the Taylor expansion and integrating with respect to the interaction times
\begin{gather}
    \mathcal{X}_2 
    =\sum_{n_L,n_R=0}^\infty (-i|\eta|)^{2(n_L+n_R+1)}  e^{h_M} \nonumber\\ \sum_{{\bf k}} \chi_M \, 
    f_{M_L,{\bf q}_L,1} (T_L)\, f_{M_R,{\bf q}_R,2} (T_R)\, 
    e^{-i (m_C \omega +\omega_3)T_C} .
\end{gather}

The two expressions above capture all the cases that are relevant for the third-order response functions, which can be obtained by suitably defining the times $T_L$, $T_C$, and $T_R$ in terms of the waiting times $T_1$, $T_2$, and $T_3$ and exploiting the fact that $U^\dagger (t)=U(-t)$.

\subsubsection{Ground state bleaching}
The rephasing component of the ground state bleaching contribution is associated to the quantity:
\begin{gather}
    \langle 1,0 | U^\dagger_e(T_1) | 1\rangle \langle 0 | U^\dagger_g(T_2+T_3) | 0 \rangle \langle 1 | U_e(T_3) | 1,0 \rangle .
\end{gather}
This can be reduced to the function $\mathcal{X}_1$ by setting: $T_L=-T_1$, $T_C=-T_2-T_3$, $T_R=T_3$.

The non-rephasing component of the ground-state bleaching contribution is associated to the quantity:
\begin{gather}
    \langle 1,0 | U_e(T_3) | 1\rangle \langle 0 | U_g(T_2) | 0 \rangle \langle 1 | U_e(T_1) | 1,0 \rangle .
\end{gather}
This can be reduced to the function $\mathcal{X}_1$ by setting: $T_L=T_3$, $T_C=T_2$, $T_R=T_1$.

\subsubsection{Stimulated emission}
The rephasing component of the stimulated emission contribution is related to the function:
\begin{gather}
    \langle 1,0 | U_e^\dagger(T_1+T_2) | 1\rangle \langle 0 | U_g^\dagger(T_3) | 0 \rangle \langle 1 | U_e(T_2+T_3) | 1,0 \rangle .
\end{gather}
This can be reduced to the quantity $\mathcal{X}_1$ by setting: $T_L=-T_1-T_2$, $T_C=-T_3$, $T_R=T_2+T_3$.

The non-rephasing component of the stimulated emission contribution is related to the function:
\begin{align}
    \langle 1,0 | U_e^\dagger(T_2) | 1\rangle \langle 0 | U_g^\dagger(T_3) | 0 \rangle \langle 1 | U_e(T_1+T_2+T_3) | 1,0 \rangle .
\end{align}
This can be reduced to the quantity $\mathcal{X}_1$ by setting: $T_L=-T_2$, $T_C=-T_3$, $T_R=T_1+T_2+T_3$.

\subsubsection{Excited state absorption}
The rephasing component of the excited state absorption is associated to the quantity:
\begin{align}
    \langle 1,0 | U_e^\dagger(T_1+T_2+T_3) | 2\rangle \langle 3 | U_b (T_3) | 3 \rangle \langle 2 | U_e(T_2) | 1,0 \rangle .
\end{align}
This can be reduced to the function $\mathcal{X}_2$ by setting: $T_L=-T_1-T_2-T_3$, $T_C=T_3$, $T_R=T_2$.

The non-rephasing component of the excited state absorption is associated to the quantity:
\begin{align}
    \langle 1,0 | U_e^\dagger(T_2+T_3) | 2\rangle \langle 3 | U_b (T_3) | 3 \rangle \langle 2 | U_e(T_1+T_2) | 1,0 \rangle
\end{align}
This can be reduced to the function $\mathcal{X}_2$ by setting: $T_L=-T_2-T_3$, $T_C=T_3$, $T_R=T_1+T_2$.

\subsubsection{Double quantum coherence}
The first component of double quantum coherence contribution is related to the function:
\begin{align}
    \langle 1,0 | U_e^\dagger(T_3) | 2\rangle \langle 3 | U_b(T_2+T_3) | 3 \rangle \langle 2 | U_e(T_1) | 1,0 \rangle
\end{align}
This can be reduced to the quantity $\mathcal{X}_2$ by setting: $T_L=-T_3$, $T_C=T_2+T_3$, $T_R=T_1$.

The second component of double quantum coherence contribution is related to the function:
\begin{align}
    \langle 1,0 | U_e(T_3) | 2\rangle \langle 3 | U_b(T_2) | 3 \rangle \langle 2 | U_e(T_1) | 1,0 \rangle
\end{align}
This can be reduced to the quantity $\mathcal{X}_2$ by setting: $T_L=T_3$, $T_C=T_2$, $T_R=T_1$.

\section{Conclusions}

In conclusion, we have developed an approach for analytically deriving the response functions $\mathcal{R}$ in model systems that include non-adiabatic couplings. The approach is based on the perturbative expansion of the relevant propagators with respect to the non-adiabatic term in the Hamiltonian, and on the formal correspondence between the contributions in the expansion and adiabatic response functions $R$, recently derived for the displaced oscillator model. {\color{black} After performing the Taylor expansion of $R$ with respect to the displacements and integrating with respect to the interaction times, we derive analytical expressions for the one- and three-time propagators and, from these, the linear and nonlinear response functions. It has also been shown that the effect of a simple and yet relevant form of decoherence, including both dephasing and relaxation, can be accounted by multiplying the above quantities by suitable exponential decay functions.}

The approach has been applied to two prototypical model systems, which have been used for modeling a number of physical systems of interest. In these cases, the response functions have been compared with those obtained by an independent numerical approach, showing the convergence of the perturbative approach for time intervals of increasing duration, as the number of terms in the expansion increases. {\color{black} General criteria are given for the convergence of the Dyson expansion, both in the time and in the frequency domains.}

The application of the present approach to higher-order response functions or to more complex models, which include more vibrational modes, electronic levels, allowed optical transitions, or more non-adiabatic terms in the Hamiltonian, is conceptually straightforward. {\color{black} In fact, it mainly requires to apply the above procedure to a number of additional pathways, that such extensions would allow. Other generalizations can also be envisaged, resulting from a different expression of the non-adiabatic term $V$ in the Hamiltonian. In particular, expressions of such term that are proportional to the nuclear position operator are often encountered in the literature. This would require an analogous generalization of the adiabatic response function to the case of nuclear-position dependent transitions amplitudes (from Franck-Condon to Herzberg-Teller coupling), which is the object of ongoing investigations.}

\acknowledgements

The author acknowledges fruitful discussions with Frank Ernesto Quintela Rodriguez.

\appendix

{\color{black}

\section{Equivalent expressions \\ of the Hamiltonian \label{app:x}}

Within the subspace $\mathcal{S}_e$, the Hamiltonian $H$ given in Eqs. (1-2) can be written as the sum of a term ($\hbar=1$)
\begin{gather}
    H_a = \alpha {\bf n} \cdot {\bf \sigma} 
    + \beta \sigma_z (a^\dagger+a) \equiv H_{a,1} + H_{a,2}
\end{gather}
and of a term $H_b$ that is proportional to the identity operator $\mathcal{I}=| 1 \rangle\langle 1 |+| 2 \rangle\langle 2 |$, and plays no role in the following discussion.
The components of ${\bf\sigma}$ are the Pauli matrices $\sigma_X$, $\sigma_Y$, and $\sigma_Z$ in the basis $\{|1\rangle,|2\rangle\}$.
The electronic part of the Hamiltonian, $H_{a,1}$, is characterized by the real coupling constant $\alpha$ and by the unit vector ${\bf n} =(\sin\theta\cos\phi,\sin\theta\sin\phi,\cos\theta)$.
The same vector can also be expressed as a function of the Hamiltonian parameters in Eqs. (\ref{eq:ham1}-\ref{eq:ham2}):
\begin{gather}
    {\bf n} = C \left[{\rm Re}(\eta),-{\rm Im}(\eta),\frac{1}{2}(\bar\omega_1-\bar\omega_2)\right],
\end{gather}
with $C=[|\eta|^2+\tfrac{1}{4}(\bar\omega_1-\bar\omega_2)^2]^{-1/2}$.

This determines the eigenstates $|+\rangle$ and $|-\rangle$ of the electronic part, which can also be written as 
\begin{gather}
    H_{a,1}= \alpha {\bf n} \cdot {\bf \sigma} = \alpha (| + \rangle\langle + | - | - \rangle\langle - |) \equiv \alpha \tau_Z .
\end{gather}
With respect to the basis $\{|+\rangle,|-\rangle\}$ and to the corresponding Pauli matrices $\tau_X$, $\tau_Y$, and $\tau_Z$, the Hamiltonian reads:
\begin{gather}
    H_a = \alpha\tau_Z + \beta{\bf m} \cdot {\bf \tau} (a^\dagger+a) ,
\end{gather}
where
${\bf m}=(-\sin\theta\cos\phi,-\sin\theta\sin\phi,\cos\theta)$, or equivalently 
\begin{gather}
    {\bf m} = C \left[-{\rm Re}(\eta),{\rm Im}(\eta),\frac{1}{2}(\bar\omega_1-\bar\omega_2)\right].
\end{gather}
Therefore, the kind of non-adiabatic Hamiltonian considered in the present paper, characterized by a transverse electronic term $V$ and an electron-vibrational coupling that is diagonal in the diabatic state basis, can also be written as the sum of a diagonal electronic term and of a more general, non-diagonal vibronic coupling.}

\section{List of the functions $f_{M,{\bf q},\sigma}(t)$ for $M\le 6$\label{app:lof}}

We consider for simplicity the case where the vibrational frequency $\omega$ and that corresponding to the electronic gap ($\bar\omega_{12}$) are incommensurate. Therefore, in view of Eq. (\ref{eq:mfr}), only the frequencies $\omega_{ij}$ where $j-i$ is an odd number can vanish. In particular, this happens if, in addition, $\sum_{k=i}^j m_k=0$. 

In the following we report, for each value of $M$: the expressions of the functions $A_j$ that apply if all the relevant frequencies $\omega_{ij}$ are nonzero; the expressions that change with respect to the above in case some of the frequencies vanish. If two frequencies $\omega_{ij}$ and $\omega_{mn}$, with $j \neq n$, vanish at the same time, the resulting changes in the functions $A_{j}$, with respect to the case where no frequencies vanish, are all the ones that are derived for $\omega_{ij}=0$ and $\omega_{mn}=0$ independently. 

{\color{black} 
\subsubsection{Zero-th order ($M=1$)}
This is the contribution of lowest order in $\eta$ to the diagonal propagators $\langle\sigma;0|U_S|\sigma;0\rangle$ ($\sigma=1,2$). It is characterized by terms with 
\begin{gather}
    A_1 = 1,
\end{gather}
with corresponding frequencies $\Omega_1=\omega q_1 +\bar\omega_\sigma$.
}

\subsubsection{First order ($M=2$)}
{\color{black} This is the contribution of lowest order in $\eta$ to the non-diagonal propagators $\langle\sigma;0|U_S|3-\sigma;0\rangle$ ($\sigma=1,2$).}
From the general expressions of the coefficients $A_{j}$ [Eq. (\ref{eq:frequencies})] it follows that: 
\begin{gather}
A_{2} = \frac{1}{i\omega_{11}},\ A_{1} = -\frac{1}{i\omega_{11}} . 
\end{gather}
The frequency $\omega_{11}$ is always nonzero. {\color{black} Therefore, while considering the Fourier transforms, the first order contribution can only give rise to Lorentzian line shapes ($\hat{f}_0$), centered at the frequencies $\Omega_2=\omega q_2+\bar\omega_{3-\sigma}$ and $\Omega_1=\omega q_1+\bar\omega_{\sigma}$. 

In the absence of displacement ($z_1=z_2=0$), all the $q_j$ vanish, and the above expressions reduce to
\begin{gather*}
    A_2=\frac{1}{i\bar\omega_{12}},\ A_1=-\frac{1}{i\bar\omega_{12}} .
\end{gather*}}

\subsubsection{Second order ($M=3$)}
{\color{black} This is the contribution of lowest nonzero order in $\eta$ to the diagonal propagators.}
From the general expressions of the coefficients $A_{j}$ [Eq. (\ref{eq:frequencies})], if all the frequencies $\omega_{ij}$ are nonzero, it follows that: 
\begin{gather}
A_{3} = -\frac{1}{\omega_{12}\omega_{22}},\ 
A_{2} = \frac{1}{\omega_{11}\omega_{22}},\  
A_{1} =-\frac{1}{\omega_{11}\omega_{12}}.
\end{gather}
{\color{black} These are multiplied by terms that oscillate at the frequencies $\Omega_{2k-1}=\omega q_{2k-1}+\bar\omega_{\sigma}$ and $\Omega_{2k}=\omega q_{2k}+\bar\omega_{3-\sigma}$.}

If instead $\omega_{12}=0$, then the following expressions replace those reported above for the general case:
\begin{gather}
    A_{3} = -\frac{it}{\omega_{22}},\ 
    A_{1} = \frac{1}{\omega_{22}^2}.
\end{gather}
{\color{black} These coefficients correspond to the frequency $\Omega_1=\Omega_3$ (the equality follows from $\omega_{12}=\Omega_1-\Omega_3=0$). 
The coefficient $A_{2}$ remains unchanged.

In the absence of displacement ($z_1=z_2=0$), the above expressions for $\omega_{12}=0$ reduce to
\begin{gather*}
    A_3=\frac{it}{\bar\omega_{12}},\ A_2=-\frac{1}{\bar\omega_{12}^2},\ A_1=\frac{1}{\bar\omega_{12}^2} .
\end{gather*}}

\subsubsection{Third order ($M=4$)}
From the general expressions of the coefficients $A_{j}$, if all the frequencies $\omega_{ij}$ are nonzero, it follows that: 
\begin{gather}
A_{4} =-\frac{1}{i\omega_{13}\omega_{23}\omega_{33}},\ 
A_{3} =\frac{1}{i\omega_{12}\omega_{22}\omega_{33}} \nonumber\\ \label{eq:A01}
A_{2} =-\frac{1}{i\omega_{11}\omega_{22}\omega_{23}},\ 
A_{1} =\frac{1}{i\omega_{11}\omega_{12}\omega_{13}} .
\end{gather}
{\color{black} The corresponding frequencies $\Omega_j$ are given by the same expressions specified for the previous orders.}

If $\omega_{12}=0$ (and therefore $\Omega_1=\Omega_3$) and $\omega_{23}\neq 0$, then the following expressions replace those reported above for the general case:
\begin{gather}\label{eq:a01}
    A_{3} = \frac{t}{\omega_{22}\omega_{33}},\ 
    A_{1} = -\frac{1}{i\omega_{22}\omega_{33}} \left( \frac{1}{\omega_{22}} - \frac{1}{\omega_{33}} \right),
\end{gather}
while $A_{2}$ and $A_{4}$ remain unchanged.

If $\omega_{23}=0$ (and therefore $\Omega_2=\Omega_4$) and $\omega_{12}\neq 0$, then the following expressions replace those reported above for the general case:
\begin{gather}\label{eq:a02}
A_{4} =-\frac{t}{\omega_{11}\omega_{33}},\ 
A_{2} = \frac{1}{i\omega_{11}\omega_{33}} \left( \frac{1}{\omega_{11}} + \frac{1}{\omega_{33}} \right),
\end{gather}
while $A_{1}$ and $A_{3}$ remain unchanged. 

If both $\omega_{12}=\omega_{23}=0$ are zero, then the functions $A_{j}$ are given by the expressions in Eqs. (\ref{eq:a01}-\ref{eq:a02}). {\color{black} In the absence of displacements, these reduce to:
\begin{gather*}
    A_4=-\frac{t}{\bar\omega_{12}^2},\  
    A_3=-\frac{t}{\bar\omega_{12}^2}, \
    A_2= \frac{2}{i\bar\omega_{12}^3},\  
    A_1=-\frac{2}{i\bar\omega_{12}^3} .
\end{gather*}}

\subsubsection{Fourth order ($M=5$)}
From the general expressions of the coefficients $A_{j}$, if all the frequencies $\omega_{ij}$ are nonzero, it follows that: 
\begin{gather}
A_{5}=\frac{1}{\omega_{14}\omega_{24}\omega_{34}\omega_{44}},\ 
A_{4}=-\frac{1}{\omega_{13}\omega_{23}\omega_{33}\omega_{44}} \nonumber\\
A_{3} = \frac{1}{\omega_{12}\omega_{22}\omega_{33}\omega_{34}},\ 
A_{2} = - \frac{1}{\omega_{11}\omega_{22}\omega_{23}\omega_{24}} 
\nonumber\\ \label{eq:A02}
A_{1} = \frac{1}{\omega_{11}\omega_{12}\omega_{13}\omega_{14}}.
\end{gather}

If $\omega_{12}=0$ ($\Omega_1=\Omega_3$), then the above expressions of $A_{1}$ and $A_{3}$ are replaced by the following ones:
\begin{gather}
A_{3}=\frac{it}{\omega_{22}\omega_{33}\omega_{34}}\nonumber\\
A_{1} = -\frac{1}{\omega_{22}\omega_{33}\omega_{34}} 
\left( 
\frac{1}{\omega_{22}} -\frac{1}{\omega_{33}} -\frac{1}{\omega_{34}}
\right).
\end{gather}

If $\omega_{14}=0$ ($\Omega_1=\Omega_5$) and $\omega_{34}\neq 0$, then:
\begin{gather}
A_{5}=\frac{it}{\omega_{24}\omega_{34}\omega_{44}}\nonumber\\
A_{1} = -\frac{1}{\omega_{24}\omega_{34}\omega_{44}} 
\left( 
\frac{1}{\omega_{24}} 
+\frac{1}{\omega_{34}} 
+\frac{1}{\omega_{44}}
\right).
\end{gather}

If $\omega_{23}=0$ ($\Omega_2=\Omega_4$), then:
\begin{gather}
A_{4}=-\frac{it}{\omega_{13}\omega_{33}\omega_{44}} \nonumber\\
A_{2}=\frac{1}{\omega_{13}\omega_{33}\omega_{44}}
\left( 
\frac{1}{\omega_{13}} +\frac{1}{\omega_{33}} -\frac{1}{\omega_{44}}
\right).
\end{gather}

If $\omega_{34}=0$ ($\Omega_3=\Omega_5$) and $\omega_{14}\neq 0$, then:
\begin{gather}
A_{5}=\frac{it}{\omega_{14}\omega_{24}\omega_{44}}\nonumber\\
A_{3}=-\frac{1}{\omega_{14}\omega_{24}\omega_{44}}
\left( 
\frac{1}{\omega_{14}} +\frac{1}{\omega_{24}} +\frac{1}{\omega_{44}}
\right).
\end{gather}

If $\omega_{14}=\omega_{34}=0$ ($\Omega_1=\Omega_3=\Omega_5$):
\begin{gather}
A_{5}=-\frac{t^2}{2\omega_{24}\omega_{44}},\ 
A_{3}=-\frac{it}{\omega_{24}\omega_{44}}\left(\frac{1}{\omega_{24}}+\frac{1}{\omega_{44}}\right)\nonumber\\
A_{1}=\frac{1}{\omega_{24}\omega_{44}} \left(\frac{1}{\omega^2_{24}}+\frac{1}{\omega^2_{44}}+\frac{1}{\omega_{24}\omega_{44}}\right).
\end{gather}

{\color{black} If the oscillator doesn't undergo any displacement in the states $|1\rangle$ and $|2\rangle$ ($z_1=z_2=0$), then all the frequencies $\omega_{ij}$ with even (odd) $i$ and odd (even) $j$ vanish. The above equations reduce to:
\begin{gather*}
    A_5=-\frac{t^2}{2\bar\omega_{12}^2}, \ 
    A_4=\frac{it}{\bar\omega_{12}^3}, \ 
    A_3=\frac{2it}{\bar\omega_{12}^3}, \\
    A_2= -\frac{3}{\bar\omega_{12}^4}, \ 
    A_1=\frac{3}{\bar\omega_{12}^4} .
\end{gather*}}

\subsubsection{Fifth order ($M=6$)}
From the general expressions of the coefficients $A_{j}$, if all the frequencies $\omega_{ij}$ are nonzero, it follows that: 
\begin{gather}
A_{6}=\frac{1}{i\omega_{15}\omega_{25}\omega_{35}\omega_{45}\omega_{55}},\  
A_{5}=-\frac{1}{i\omega_{14}\omega_{24}\omega_{34}\omega_{44}\omega_{55}}\nonumber\\ 
A_{4}=\frac{1}{i\omega_{13}\omega_{23}\omega_{33}\omega_{44}\omega_{45}},\ 
A_{3}=-\frac{1}{i\omega_{12}\omega_{22}\omega_{33}\omega_{34}\omega_{35}}
\nonumber\\
A_{2}=\frac{1}{i\omega_{11}\omega_{22}\omega_{23}\omega_{24}\omega_{25}},\ 
A_{1}=-\frac{1}{i\omega_{11}\omega_{12}\omega_{13}\omega_{14}\omega_{15}}.
\end{gather}
{\color{black} The Fourier transform of these contributions thus give rise to Lorentzian line shapes ($\hat{f}_0$), centered at the frequencies $\Omega_j$.

Other vectors ${\bf q}$ in the Taylor expansion will give rise to vanishing frequencies. We start by considering the case where only one frequency vanishes within each group $\omega_{ik}$ ($i=1,\dots,k$).}
If $\omega_{12}=0$ (and therefore $\Omega_1=\Omega_3$), then:
\begin{gather}
A_{3}=-\frac{t}{\omega_{22}\omega_{33}\omega_{34}\omega_{35}}\nonumber\\
A_{1}=\frac{1}{i\omega_{22}\omega_{33}\omega_{34}\omega_{35}}
\left( 
\frac{1}{\omega_{22}} -\frac{1}{\omega_{33}} -\frac{1}{\omega_{34}}
-\frac{1}{\omega_{35}}
\right).
\end{gather}

If $\omega_{14}=0$ (and therefore $\Omega_1=\Omega_5$) and $\omega_{34}\neq 0$, then:
\begin{gather}
    A_{5}=-\frac{t}{\omega_{24}\omega_{34}\omega_{44}\omega_{55}}\nonumber\\
A_{1}=\frac{1}{i\omega_{24}\omega_{34}\omega_{44}\omega_{55}}
\left( 
\frac{1}{\omega_{24}} +\frac{1}{\omega_{34}} +\frac{1}{\omega_{44}}
-\frac{1}{\omega_{55}}
\right).
\end{gather}

If $\omega_{23}=0$ (and therefore $\Omega_2=\Omega_4$), then:
\begin{gather}
A_{4}=\frac{t}{\omega_{13}\omega_{33}\omega_{44}\omega_{45}},\ 
\nonumber\\
A_{2}=-\frac{1}{i\omega_{13}\omega_{33}\omega_{44}\omega_{45}}
\left( 
\frac{1}{\omega_{13}} +\frac{1}{\omega_{33}} -\frac{1}{\omega_{44}}
-\frac{1}{\omega_{45}}
\right) .
\end{gather}

If $\omega_{25}=0$ (and therefore $\Omega_2=\Omega_6$) and $\omega_{45}\neq 0$, then:
\begin{gather}
A_{6}=\frac{t}{\omega_{15}\omega_{35}\omega_{45}\omega_{55}}\\
A_{2}=\frac{1}{i\omega_{15}\omega_{35}\omega_{45}\omega_{55}}
\left( 
\frac{1}{\omega_{15}} +\frac{1}{\omega_{35}} +\frac{1}{\omega_{45}}
+\frac{1}{\omega_{55}}
\right) 
\end{gather}

If $\omega_{34}=0$ (and therefore $\Omega_3=\Omega_5$) and $\omega_{14}\neq 0$, then:
\begin{gather}
A_{5}=-\frac{t}{\omega_{14}\omega_{24}\omega_{44}\omega_{55}}\\ A_{3}=\frac{1}{i\omega_{14}\omega_{24}\omega_{44}\omega_{55}}
\left( 
\frac{1}{\omega_{14}} 
+\frac{1}{\omega_{24}} 
+\frac{1}{\omega_{44}}
-\frac{1}{\omega_{55}}
\right) .
\end{gather}

If $\omega_{45}=0$ (and therefore $\Omega_4=\Omega_6$) and $\omega_{25}\neq 0$, then:
\begin{gather}
A_{6}=\frac{t}{\omega_{15}\omega_{25}\omega_{35}\omega_{55}}\\
A_{4}=-\frac{1}{i\omega_{15}\omega_{25}\omega_{35}\omega_{55}}
\left( 
\frac{1}{\omega_{15}} 
+\frac{1}{\omega_{25}} 
+\frac{1}{\omega_{35}}
+\frac{1}{\omega_{55}}
\right) .
\end{gather}
{\color{black} In all these cases, the Fourier transform gives rise to functions $\hat{f}_0$ and $\hat{f}_1$, both centered at the relevant frequencies $\Omega_j$.

We finally consider the case where two frequencies vanish within each group $\omega_{ik}$ ($i=1,\dots,k$).}
If $\omega_{14}=\omega_{34}=0$ (and therefore $\Omega_1=\Omega_3=\Omega_5$):
\begin{gather}
A_{5}=\frac{t^2}{2i\omega_{24}\omega_{44}\omega_{55}}\nonumber\\
A_{3}=\frac{t}{\omega_{24}\omega_{44}\omega_{55}}\left(\frac{1}{\omega_{24}}+\frac{1}{\omega_{44}}-\frac{1}{\omega_{55}}\right)\nonumber\\
A_{1}=-\frac{1}{i\omega_{24}\omega_{44}\omega_{55}} \left(\frac{1}{\omega^2_{24}}+\frac{1}{\omega^2_{44}}+\frac{1}{\omega^2_{55}}\right.\nonumber\\ \left.+\frac{1}{\omega_{24}\omega_{44}}-\frac{1}{\omega_{24}\omega_{55}}-\frac{1}{\omega_{44}\omega_{55}}\right).
\end{gather}

If $\omega_{25}=\omega_{45}=0$ ($\Omega_2=\Omega_4=\Omega_6$):
\begin{gather}
A_{6}=-\frac{t^2}{2i\omega_{15}\omega_{35}\omega_{55}}\nonumber\\
A_{4}=-\frac{t}{\omega_{15}\omega_{35}\omega_{55}}\left(\frac{1}{\omega_{15}}+\frac{1}{\omega_{35}}+\frac{1}{\omega_{55}}\right)\nonumber\\
A_{2}=\frac{1}{i\omega_{15}\omega_{35}\omega_{55}} \left(\frac{1}{\omega^2_{15}}+\frac{1}{\omega^2_{35}}+\frac{1}{\omega^2_{55}}\right.\nonumber\\ \left.+\frac{1}{\omega_{15}\omega_{35}}+\frac{1}{\omega_{15}\omega_{55}}+\frac{1}{\omega_{35}\omega_{55}}\right).
\end{gather}

{\color{black} If the the displacements corresponding to the states $|1\rangle$ and $|2\rangle$ vanish, then $\omega_{ij}=0$ for even (odd) $i$ and odd (even) $j$. The above equations thus reduce to:
\begin{gather*}
    A_6=-\frac{t^2}{2i\bar\omega_{12}^3},\  
    A_5=\frac{t^2}{2i\bar\omega_{12}^3},\  
    A_4=-\frac{3t}{\bar\omega_{12}^4}, \\
    A_3=-\frac{3t}{\bar\omega_{12}^4},\  
    A_2=\frac{6}{i\bar\omega_{12}^5},\  
    A_1=-\frac{6}{i\bar\omega_{12}^5} .
\end{gather*}}


\begin{thebibliography}{42}%
\makeatletter
\providecommand \@ifxundefined [1]{%
 \@ifx{#1\undefined}
}%
\providecommand \@ifnum [1]{%
 \ifnum #1\expandafter \@firstoftwo
 \else \expandafter \@secondoftwo
 \fi
}%
\providecommand \@ifx [1]{%
 \ifx #1\expandafter \@firstoftwo
 \else \expandafter \@secondoftwo
 \fi
}%
\providecommand \natexlab [1]{#1}%
\providecommand \enquote  [1]{``#1''}%
\providecommand \bibnamefont  [1]{#1}%
\providecommand \bibfnamefont [1]{#1}%
\providecommand \citenamefont [1]{#1}%
\providecommand \href@noop [0]{\@secondoftwo}%
\providecommand \href [0]{\begingroup \@sanitize@url \@href}%
\providecommand \@href[1]{\@@startlink{#1}\@@href}%
\providecommand \@@href[1]{\endgroup#1\@@endlink}%
\providecommand \@sanitize@url [0]{\catcode `\\12\catcode `\$12\catcode
  `\&12\catcode `\#12\catcode `\^12\catcode `\_12\catcode `\%12\relax}%
\providecommand \@@startlink[1]{}%
\providecommand \@@endlink[0]{}%
\providecommand \url  [0]{\begingroup\@sanitize@url \@url }%
\providecommand \@url [1]{\endgroup\@href {#1}{\urlprefix }}%
\providecommand \urlprefix  [0]{URL }%
\providecommand \Eprint [0]{\href }%
\providecommand \doibase [0]{http://dx.doi.org/}%
\providecommand \selectlanguage [0]{\@gobble}%
\providecommand \bibinfo  [0]{\@secondoftwo}%
\providecommand \bibfield  [0]{\@secondoftwo}%
\providecommand \translation [1]{[#1]}%
\providecommand \BibitemOpen [0]{}%
\providecommand \bibitemStop [0]{}%
\providecommand \bibitemNoStop [0]{.\EOS\space}%
\providecommand \EOS [0]{\spacefactor3000\relax}%
\providecommand \BibitemShut  [1]{\csname bibitem#1\endcsname}%
\let\auto@bib@innerbib\@empty
\bibitem [{\citenamefont {Mukamel}(1995)}]{Mukamel95a}%
  \BibitemOpen
  \bibfield  {author} {\bibinfo {author} {\bibfnamefont {S.}~\bibnamefont
  {Mukamel}},\ }\href@noop {} {\emph {\bibinfo {title} {Principles of Nonlinear
  Optical Spectroscopy}}}\ (\bibinfo  {publisher} {Oxford University Press},\
  \bibinfo {year} {1995})\BibitemShut {NoStop}%
\bibitem [{\citenamefont {Hamm}\ and\ \citenamefont {Zanni}(2011)}]{Hamm11a}%
  \BibitemOpen
  \bibfield  {author} {\bibinfo {author} {\bibfnamefont {P.}~\bibnamefont
  {Hamm}}\ and\ \bibinfo {author} {\bibfnamefont {M.~T.}\ \bibnamefont
  {Zanni}},\ }\href {https://doi.org/10.1017/CBO9780511675935} {\emph {\bibinfo
  {title} {Concepts and Methods of 2D Infrared Spectroscopy}}}\ (\bibinfo
  {publisher} {Cambridge University Press},\ \bibinfo {year}
  {2011})\BibitemShut {NoStop}%
\bibitem [{\citenamefont {Scholes}\ \emph {et~al.}(2017)\citenamefont
  {Scholes}, \citenamefont {Fleming}, \citenamefont {Chen}, \citenamefont
  {Aspuru-Guzik}, \citenamefont {Buchleitner}, \citenamefont {Coker},
  \citenamefont {Engel}, \citenamefont {van Grondelle}, \citenamefont
  {Ishizaki},\ and\ \citenamefont {Jonas}}]{Scholes2017}%
  \BibitemOpen
  \bibfield  {author} {\bibinfo {author} {\bibfnamefont {G.~D.}\ \bibnamefont
  {Scholes}}, \bibinfo {author} {\bibfnamefont {G.~R.}\ \bibnamefont
  {Fleming}}, \bibinfo {author} {\bibfnamefont {L.~X.}\ \bibnamefont {Chen}},
  \bibinfo {author} {\bibfnamefont {A.}~\bibnamefont {Aspuru-Guzik}}, \bibinfo
  {author} {\bibfnamefont {A.}~\bibnamefont {Buchleitner}}, \bibinfo {author}
  {\bibfnamefont {D.~F.}\ \bibnamefont {Coker}}, \bibinfo {author}
  {\bibfnamefont {G.~S.}\ \bibnamefont {Engel}}, \bibinfo {author}
  {\bibfnamefont {R.}~\bibnamefont {van Grondelle}}, \bibinfo {author}
  {\bibfnamefont {A.}~\bibnamefont {Ishizaki}}, \ and\ \bibinfo {author}
  {\bibfnamefont {D.~M.}\ \bibnamefont {Jonas}},\ }\href
  {https://doi.org/10.1038/nature21425} {\bibfield  {journal} {\bibinfo
  {journal} {Nature}\ }\textbf {\bibinfo {volume} {543}},\ \bibinfo {pages}
  {647} (\bibinfo {year} {2017})}\BibitemShut {NoStop}%
\bibitem [{\citenamefont {Smallwood}\ and\ \citenamefont
  {Cundiff}(2018)}]{Smallwood18a}%
  \BibitemOpen
  \bibfield  {author} {\bibinfo {author} {\bibfnamefont {C.~L.}\ \bibnamefont
  {Smallwood}}\ and\ \bibinfo {author} {\bibfnamefont {S.~T.}\ \bibnamefont
  {Cundiff}},\ }\href {\doibase https://doi.org/10.1002/lpor.201800171}
  {\bibfield  {journal} {\bibinfo  {journal} {Laser \& Photonics Reviews}\
  }\textbf {\bibinfo {volume} {12}},\ \bibinfo {pages} {1800171} (\bibinfo
  {year} {2018})}\BibitemShut {NoStop}%
\bibitem [{\citenamefont {Rozzi}\ \emph {et~al.}(2018)\citenamefont {Rozzi},
  \citenamefont {Troiani},\ and\ \citenamefont {Tavernelli}}]{Rozzi18a}%
  \BibitemOpen
  \bibfield  {author} {\bibinfo {author} {\bibfnamefont {C.~A.}\ \bibnamefont
  {Rozzi}}, \bibinfo {author} {\bibfnamefont {F.}~\bibnamefont {Troiani}}, \
  and\ \bibinfo {author} {\bibfnamefont {I.}~\bibnamefont {Tavernelli}},\
  }\href {\doibase 10.1088/1361-648X/aa948a} {\bibfield  {journal} {\bibinfo
  {journal} {Journal of Physics: Condensed Matter}\ }\textbf {\bibinfo {volume}
  {30}},\ \bibinfo {pages} {013002} (\bibinfo {year} {2018})}\BibitemShut
  {NoStop}%
\bibitem [{\citenamefont {Collini}(2021)}]{Collini2021}%
  \BibitemOpen
  \bibfield  {author} {\bibinfo {author} {\bibfnamefont {E.}~\bibnamefont
  {Collini}},\ }\href {https://doi.org/10.1021/acs.jpcc.1c02693} {\bibfield
  {journal} {\bibinfo  {journal} {J. Phys. Chem. C}\ }\textbf {\bibinfo
  {volume} {125}},\ \bibinfo {pages} {13096} (\bibinfo {year}
  {2021})}\BibitemShut {NoStop}%
\bibitem [{\citenamefont {Chin}\ \emph {et~al.}(2013)\citenamefont {Chin},
  \citenamefont {Prior}, \citenamefont {Rosenbach}, \citenamefont
  {Caycedo-Soler}, \citenamefont {Huelga},\ and\ \citenamefont
  {Plenio}}]{Chin2013}%
  \BibitemOpen
  \bibfield  {author} {\bibinfo {author} {\bibfnamefont {A.~W.}\ \bibnamefont
  {Chin}}, \bibinfo {author} {\bibfnamefont {J.}~\bibnamefont {Prior}},
  \bibinfo {author} {\bibfnamefont {R.}~\bibnamefont {Rosenbach}}, \bibinfo
  {author} {\bibfnamefont {F.}~\bibnamefont {Caycedo-Soler}}, \bibinfo {author}
  {\bibfnamefont {S.~F.}\ \bibnamefont {Huelga}}, \ and\ \bibinfo {author}
  {\bibfnamefont {M.~B.}\ \bibnamefont {Plenio}},\ }\href {\doibase
  10.1038/nphys2515} {\bibfield  {journal} {\bibinfo  {journal} {Nature
  Physics}\ }\textbf {\bibinfo {volume} {9}},\ \bibinfo {pages} {113} (\bibinfo
  {year} {2013})}\BibitemShut {NoStop}%
\bibitem [{\citenamefont {Falke}\ \emph {et~al.}(2014)\citenamefont {Falke},
  \citenamefont {Rozzi}, \citenamefont {Brida}, \citenamefont {Maiuri},
  \citenamefont {Amato}, \citenamefont {Sommer}, \citenamefont {Sio},
  \citenamefont {Rubio}, \citenamefont {Cerullo}, \citenamefont {Molinari},\
  and\ \citenamefont {Lienau}}]{Falke14a}%
  \BibitemOpen
  \bibfield  {author} {\bibinfo {author} {\bibfnamefont {S.~M.}\ \bibnamefont
  {Falke}}, \bibinfo {author} {\bibfnamefont {C.~A.}\ \bibnamefont {Rozzi}},
  \bibinfo {author} {\bibfnamefont {D.}~\bibnamefont {Brida}}, \bibinfo
  {author} {\bibfnamefont {M.}~\bibnamefont {Maiuri}}, \bibinfo {author}
  {\bibfnamefont {M.}~\bibnamefont {Amato}}, \bibinfo {author} {\bibfnamefont
  {E.}~\bibnamefont {Sommer}}, \bibinfo {author} {\bibfnamefont {A.~D.}\
  \bibnamefont {Sio}}, \bibinfo {author} {\bibfnamefont {A.}~\bibnamefont
  {Rubio}}, \bibinfo {author} {\bibfnamefont {G.}~\bibnamefont {Cerullo}},
  \bibinfo {author} {\bibfnamefont {E.}~\bibnamefont {Molinari}}, \ and\
  \bibinfo {author} {\bibfnamefont {C.}~\bibnamefont {Lienau}},\ }\href
  {\doibase 10.1126/science.1249771} {\bibfield  {journal} {\bibinfo  {journal}
  {Science}\ }\textbf {\bibinfo {volume} {344}},\ \bibinfo {pages} {1001}
  (\bibinfo {year} {2014})}\BibitemShut {NoStop}%
\bibitem [{\citenamefont {Romero}\ \emph {et~al.}(2014)\citenamefont {Romero},
  \citenamefont {Augulis}, \citenamefont {Novoderezhkin}, \citenamefont
  {Ferretti}, \citenamefont {Thieme}, \citenamefont {Zigmantas},\ and\
  \citenamefont {van Grondelle}}]{Romero2014}%
  \BibitemOpen
  \bibfield  {author} {\bibinfo {author} {\bibfnamefont {E.}~\bibnamefont
  {Romero}}, \bibinfo {author} {\bibfnamefont {R.}~\bibnamefont {Augulis}},
  \bibinfo {author} {\bibfnamefont {V.~I.}\ \bibnamefont {Novoderezhkin}},
  \bibinfo {author} {\bibfnamefont {M.}~\bibnamefont {Ferretti}}, \bibinfo
  {author} {\bibfnamefont {J.}~\bibnamefont {Thieme}}, \bibinfo {author}
  {\bibfnamefont {D.}~\bibnamefont {Zigmantas}}, \ and\ \bibinfo {author}
  {\bibfnamefont {R.}~\bibnamefont {van Grondelle}},\ }\href {\doibase
  10.1038/nphys3017} {\bibfield  {journal} {\bibinfo  {journal} {Nature
  Physics}\ }\textbf {\bibinfo {volume} {10}},\ \bibinfo {pages} {676}
  (\bibinfo {year} {2014})}\BibitemShut {NoStop}%
\bibitem [{\citenamefont {O'Reilly}\ and\ \citenamefont
  {Olaya-Castro}(2014)}]{OReilly2014}%
  \BibitemOpen
  \bibfield  {author} {\bibinfo {author} {\bibfnamefont {E.~J.}\ \bibnamefont
  {O'Reilly}}\ and\ \bibinfo {author} {\bibfnamefont {A.}~\bibnamefont
  {Olaya-Castro}},\ }\href {\doibase 10.1038/ncomms4012} {\bibfield  {journal}
  {\bibinfo  {journal} {Nature Communications}\ }\textbf {\bibinfo {volume}
  {5}},\ \bibinfo {pages} {3012} (\bibinfo {year} {2014})}\BibitemShut
  {NoStop}%
\bibitem [{\citenamefont {De~Sio}\ \emph {et~al.}(2016)\citenamefont {De~Sio},
  \citenamefont {Troiani}, \citenamefont {Maiuri}, \citenamefont {R{\'e}hault},
  \citenamefont {Sommer}, \citenamefont {Lim}, \citenamefont {Huelga},
  \citenamefont {Plenio}, \citenamefont {Rozzi}, \citenamefont {Cerullo},
  \citenamefont {Molinari},\ and\ \citenamefont {Lienau}}]{DeSio16a}%
  \BibitemOpen
  \bibfield  {author} {\bibinfo {author} {\bibfnamefont {A.}~\bibnamefont
  {De~Sio}}, \bibinfo {author} {\bibfnamefont {F.}~\bibnamefont {Troiani}},
  \bibinfo {author} {\bibfnamefont {M.}~\bibnamefont {Maiuri}}, \bibinfo
  {author} {\bibfnamefont {J.}~\bibnamefont {R{\'e}hault}}, \bibinfo {author}
  {\bibfnamefont {E.}~\bibnamefont {Sommer}}, \bibinfo {author} {\bibfnamefont
  {J.}~\bibnamefont {Lim}}, \bibinfo {author} {\bibfnamefont {S.~F.}\
  \bibnamefont {Huelga}}, \bibinfo {author} {\bibfnamefont {M.~B.}\
  \bibnamefont {Plenio}}, \bibinfo {author} {\bibfnamefont {C.~A.}\
  \bibnamefont {Rozzi}}, \bibinfo {author} {\bibfnamefont {G.}~\bibnamefont
  {Cerullo}}, \bibinfo {author} {\bibfnamefont {E.}~\bibnamefont {Molinari}}, \
  and\ \bibinfo {author} {\bibfnamefont {C.}~\bibnamefont {Lienau}},\ }\href
  {\doibase 10.1038/ncomms13742} {\bibfield  {journal} {\bibinfo  {journal}
  {Nature Communications}\ }\textbf {\bibinfo {volume} {7}},\ \bibinfo {pages}
  {13742} (\bibinfo {year} {2016})}\BibitemShut {NoStop}%
\bibitem [{\citenamefont {Thouin}\ \emph {et~al.}(2019)\citenamefont {Thouin},
  \citenamefont {Valverde-Ch{\'a}vez}, \citenamefont {Quarti}, \citenamefont
  {Cortecchia}, \citenamefont {Bargigia}, \citenamefont {Beljonne},
  \citenamefont {Petrozza}, \citenamefont {Silva},\ and\ \citenamefont
  {Srimath~Kandada}}]{Thouin2019}%
  \BibitemOpen
  \bibfield  {author} {\bibinfo {author} {\bibfnamefont {F.}~\bibnamefont
  {Thouin}}, \bibinfo {author} {\bibfnamefont {D.~A.}\ \bibnamefont
  {Valverde-Ch{\'a}vez}}, \bibinfo {author} {\bibfnamefont {C.}~\bibnamefont
  {Quarti}}, \bibinfo {author} {\bibfnamefont {D.}~\bibnamefont {Cortecchia}},
  \bibinfo {author} {\bibfnamefont {I.}~\bibnamefont {Bargigia}}, \bibinfo
  {author} {\bibfnamefont {D.}~\bibnamefont {Beljonne}}, \bibinfo {author}
  {\bibfnamefont {A.}~\bibnamefont {Petrozza}}, \bibinfo {author}
  {\bibfnamefont {C.}~\bibnamefont {Silva}}, \ and\ \bibinfo {author}
  {\bibfnamefont {A.~R.}\ \bibnamefont {Srimath~Kandada}},\ }\href
  {https://doi.org/10.1038/s41563-018-0262-7} {\bibfield  {journal} {\bibinfo
  {journal} {Nat. Mater.}\ }\textbf {\bibinfo {volume} {18}},\ \bibinfo {pages}
  {349} (\bibinfo {year} {2019})}\BibitemShut {NoStop}%
\bibitem [{\citenamefont {Rafiq}\ \emph {et~al.}(2021)\citenamefont {Rafiq},
  \citenamefont {Fu}, \citenamefont {Kudisch},\ and\ \citenamefont
  {Scholes}}]{Rafiq2021}%
  \BibitemOpen
  \bibfield  {author} {\bibinfo {author} {\bibfnamefont {S.}~\bibnamefont
  {Rafiq}}, \bibinfo {author} {\bibfnamefont {B.}~\bibnamefont {Fu}}, \bibinfo
  {author} {\bibfnamefont {B.}~\bibnamefont {Kudisch}}, \ and\ \bibinfo
  {author} {\bibfnamefont {G.~D.}\ \bibnamefont {Scholes}},\ }\href
  {https://doi.org/10.1038/s41557-020-00607-9} {\bibfield  {journal} {\bibinfo
  {journal} {Nat. Chem.}\ }\textbf {\bibinfo {volume} {13}},\ \bibinfo {pages}
  {70} (\bibinfo {year} {2021})}\BibitemShut {NoStop}%
\bibitem [{\citenamefont {Kumar}\ \emph {et~al.}(2001)\citenamefont {Kumar},
  \citenamefont {Rosca}, \citenamefont {Widom},\ and\ \citenamefont
  {Champion}}]{Kumar01a}%
  \BibitemOpen
  \bibfield  {author} {\bibinfo {author} {\bibfnamefont {A.~T.~N.}\
  \bibnamefont {Kumar}}, \bibinfo {author} {\bibfnamefont {F.}~\bibnamefont
  {Rosca}}, \bibinfo {author} {\bibfnamefont {A.}~\bibnamefont {Widom}}, \ and\
  \bibinfo {author} {\bibfnamefont {P.~M.}\ \bibnamefont {Champion}},\ }\href
  {\doibase 10.1063/1.1329640} {\bibfield  {journal} {\bibinfo  {journal} {The
  Journal of Chemical Physics}\ }\textbf {\bibinfo {volume} {114}},\ \bibinfo
  {pages} {701} (\bibinfo {year} {2001})}\BibitemShut {NoStop}%
\bibitem [{\citenamefont {Egorova}\ \emph {et~al.}(2007)\citenamefont
  {Egorova}, \citenamefont {Gelin},\ and\ \citenamefont
  {Domcke}}]{Egorova2007a}%
  \BibitemOpen
  \bibfield  {author} {\bibinfo {author} {\bibfnamefont {D.}~\bibnamefont
  {Egorova}}, \bibinfo {author} {\bibfnamefont {M.~F.}\ \bibnamefont {Gelin}},
  \ and\ \bibinfo {author} {\bibfnamefont {W.}~\bibnamefont {Domcke}},\ }\href
  {\doibase 10.1063/1.2435353} {\bibfield  {journal} {\bibinfo  {journal} {The
  Journal of Chemical Physics}\ }\textbf {\bibinfo {volume} {126}},\ \bibinfo
  {pages} {074314} (\bibinfo {year} {2007})}\BibitemShut {NoStop}%
\bibitem [{\citenamefont {Mančal}\ \emph {et~al.}(2010)\citenamefont
  {Mančal}, \citenamefont {Nemeth}, \citenamefont {Milota}, \citenamefont
  {Lukeš}, \citenamefont {Kauffmann},\ and\ \citenamefont
  {Sperling}}]{Mancal10a}%
  \BibitemOpen
  \bibfield  {author} {\bibinfo {author} {\bibfnamefont {T.}~\bibnamefont
  {Mančal}}, \bibinfo {author} {\bibfnamefont {A.}~\bibnamefont {Nemeth}},
  \bibinfo {author} {\bibfnamefont {F.}~\bibnamefont {Milota}}, \bibinfo
  {author} {\bibfnamefont {V.}~\bibnamefont {Lukeš}}, \bibinfo {author}
  {\bibfnamefont {H.~F.}\ \bibnamefont {Kauffmann}}, \ and\ \bibinfo {author}
  {\bibfnamefont {J.}~\bibnamefont {Sperling}},\ }\href {\doibase
  10.1063/1.3404405} {\bibfield  {journal} {\bibinfo  {journal} {The Journal of
  Chemical Physics}\ }\textbf {\bibinfo {volume} {132}},\ \bibinfo {pages}
  {184515} (\bibinfo {year} {2010})}\BibitemShut {NoStop}%
\bibitem [{\citenamefont {Pollard}\ \emph {et~al.}(1990)\citenamefont
  {Pollard}, \citenamefont {Lee},\ and\ \citenamefont
  {Mathies}}]{Pollard1990a}%
  \BibitemOpen
  \bibfield  {author} {\bibinfo {author} {\bibfnamefont {W.~T.}\ \bibnamefont
  {Pollard}}, \bibinfo {author} {\bibfnamefont {S.~Y.}\ \bibnamefont {Lee}}, \
  and\ \bibinfo {author} {\bibfnamefont {R.~A.}\ \bibnamefont {Mathies}},\
  }\href {https://doi.org/10.1063/1.457815} {\bibfield  {journal} {\bibinfo
  {journal} {J. Chem. Phys.}\ }\textbf {\bibinfo {volume} {92}},\ \bibinfo
  {pages} {4012} (\bibinfo {year} {1990})}\BibitemShut {NoStop}%
\bibitem [{\citenamefont {Pollard}\ \emph {et~al.}(1992)\citenamefont
  {Pollard}, \citenamefont {Dexheimer}, \citenamefont {Wang}, \citenamefont
  {Peteanu}, \citenamefont {Shank},\ and\ \citenamefont
  {Mathies}}]{Pollard1992a}%
  \BibitemOpen
  \bibfield  {author} {\bibinfo {author} {\bibfnamefont {W.~T.}\ \bibnamefont
  {Pollard}}, \bibinfo {author} {\bibfnamefont {S.~L.}\ \bibnamefont
  {Dexheimer}}, \bibinfo {author} {\bibfnamefont {Q.}~\bibnamefont {Wang}},
  \bibinfo {author} {\bibfnamefont {L.~A.}\ \bibnamefont {Peteanu}}, \bibinfo
  {author} {\bibfnamefont {C.~V.}\ \bibnamefont {Shank}}, \ and\ \bibinfo
  {author} {\bibfnamefont {R.~A.}\ \bibnamefont {Mathies}},\ }\href
  {https://doi.org/10.1021/j100194a013} {\bibfield  {journal} {\bibinfo
  {journal} {J. Phys. Chem.}\ }\textbf {\bibinfo {volume} {96}},\ \bibinfo
  {pages} {6147} (\bibinfo {year} {1992})}\BibitemShut {NoStop}%
\bibitem [{\citenamefont {Butkus}\ \emph {et~al.}(2012)\citenamefont {Butkus},
  \citenamefont {Zigmantas}, \citenamefont {Valkunas},\ and\ \citenamefont
  {Abramavicius}}]{Butkus12a}%
  \BibitemOpen
  \bibfield  {author} {\bibinfo {author} {\bibfnamefont {V.}~\bibnamefont
  {Butkus}}, \bibinfo {author} {\bibfnamefont {D.}~\bibnamefont {Zigmantas}},
  \bibinfo {author} {\bibfnamefont {L.}~\bibnamefont {Valkunas}}, \ and\
  \bibinfo {author} {\bibfnamefont {D.}~\bibnamefont {Abramavicius}},\ }\href
  {\doibase https://doi.org/10.1016/j.cplett.2012.07.014} {\bibfield  {journal}
  {\bibinfo  {journal} {Chemical Physics Letters}\ }\textbf {\bibinfo {volume}
  {545}},\ \bibinfo {pages} {40} (\bibinfo {year} {2012})}\BibitemShut
  {NoStop}%
\bibitem [{\citenamefont {Cina}\ \emph {et~al.}(2016)\citenamefont {Cina},
  \citenamefont {Kovac}, \citenamefont {Jumper}, \citenamefont {Dean},\ and\
  \citenamefont {Scholes}}]{Cina2016}%
  \BibitemOpen
  \bibfield  {author} {\bibinfo {author} {\bibfnamefont {J.~A.}\ \bibnamefont
  {Cina}}, \bibinfo {author} {\bibfnamefont {P.~A.}\ \bibnamefont {Kovac}},
  \bibinfo {author} {\bibfnamefont {C.~C.}\ \bibnamefont {Jumper}}, \bibinfo
  {author} {\bibfnamefont {J.~C.}\ \bibnamefont {Dean}}, \ and\ \bibinfo
  {author} {\bibfnamefont {G.~D.}\ \bibnamefont {Scholes}},\ }\href
  {https://doi.org/10.1063/1.4947568} {\bibfield  {journal} {\bibinfo
  {journal} {J. Chem. Phys.}\ }\textbf {\bibinfo {volume} {144}},\ \bibinfo
  {pages} {175102} (\bibinfo {year} {2016})}\BibitemShut {NoStop}%
\bibitem [{\citenamefont {Le}\ \emph {et~al.}(2021)\citenamefont {Le},
  \citenamefont {Leng},\ and\ \citenamefont {Tan}}]{Le21a}%
  \BibitemOpen
  \bibfield  {author} {\bibinfo {author} {\bibfnamefont {D.~V.}\ \bibnamefont
  {Le}}, \bibinfo {author} {\bibfnamefont {X.}~\bibnamefont {Leng}}, \ and\
  \bibinfo {author} {\bibfnamefont {H.-S.}\ \bibnamefont {Tan}},\ }\href
  {\doibase https://doi.org/10.1016/j.chemphys.2021.111142} {\bibfield
  {journal} {\bibinfo  {journal} {Chemical Physics}\ }\textbf {\bibinfo
  {volume} {546}},\ \bibinfo {pages} {111142} (\bibinfo {year}
  {2021})}\BibitemShut {NoStop}%
\bibitem [{\citenamefont {Turner}\ and\ \citenamefont
  {Arpin}(2020)}]{Turner2020}%
  \BibitemOpen
  \bibfield  {author} {\bibinfo {author} {\bibfnamefont {D.~B.}\ \bibnamefont
  {Turner}}\ and\ \bibinfo {author} {\bibfnamefont {P.~C.}\ \bibnamefont
  {Arpin}},\ }\href {https://doi.org/10.1016/j.chemphys.2020.110948} {\bibfield
   {journal} {\bibinfo  {journal} {Chem. Phys.}\ }\textbf {\bibinfo {volume}
  {539}},\ \bibinfo {pages} {110948} (\bibinfo {year} {2020})}\BibitemShut
  {NoStop}%
\bibitem [{\citenamefont {Quintela~Rodriguez}\ and\ \citenamefont
  {Troiani}(2022)}]{Quintela2022a}%
  \BibitemOpen
  \bibfield  {author} {\bibinfo {author} {\bibfnamefont {F.~E.}\ \bibnamefont
  {Quintela~Rodriguez}}\ and\ \bibinfo {author} {\bibfnamefont
  {F.}~\bibnamefont {Troiani}},\ }\href {\doibase 10.1063/5.0094512} {\bibfield
   {journal} {\bibinfo  {journal} {The Journal of Chemical Physics}\ }\textbf
  {\bibinfo {volume} {157}},\ \bibinfo {pages} {034107} (\bibinfo {year}
  {2022})}\BibitemShut {NoStop}%
\bibitem [{\citenamefont {Park}\ and\ \citenamefont {Cho}(2000)}]{Park2000a}%
  \BibitemOpen
  \bibfield  {author} {\bibinfo {author} {\bibfnamefont {K.}~\bibnamefont
  {Park}}\ and\ \bibinfo {author} {\bibfnamefont {M.}~\bibnamefont {Cho}},\
  }\href {\doibase 10.1063/1.481684} {\bibfield  {journal} {\bibinfo  {journal}
  {The Journal of Chemical Physics}\ }\textbf {\bibinfo {volume} {112}},\
  \bibinfo {pages} {10496} (\bibinfo {year} {2000})}\BibitemShut {NoStop}%
\bibitem [{\citenamefont {Arpin}\ and\ \citenamefont
  {Turner}(2021)}]{Arpin21a}%
  \BibitemOpen
  \bibfield  {author} {\bibinfo {author} {\bibfnamefont {P.~C.}\ \bibnamefont
  {Arpin}}\ and\ \bibinfo {author} {\bibfnamefont {D.~B.}\ \bibnamefont
  {Turner}},\ }\href {\doibase 10.1021/acs.jpca.0c10807} {\bibfield  {journal}
  {\bibinfo  {journal} {The Journal of Physical Chemistry A}\ }\textbf
  {\bibinfo {volume} {125}},\ \bibinfo {pages} {2425} (\bibinfo {year}
  {2021})}\BibitemShut {NoStop}%
\bibitem [{\citenamefont {Schultz}\ \emph {et~al.}(2022)\citenamefont
  {Schultz}, \citenamefont {Kim}, \citenamefont {O'Connor}, \citenamefont
  {Young},\ and\ \citenamefont {Wasielewski}}]{Schultz22a}%
  \BibitemOpen
  \bibfield  {author} {\bibinfo {author} {\bibfnamefont {J.~D.}\ \bibnamefont
  {Schultz}}, \bibinfo {author} {\bibfnamefont {T.}~\bibnamefont {Kim}},
  \bibinfo {author} {\bibfnamefont {J.~P.}\ \bibnamefont {O'Connor}}, \bibinfo
  {author} {\bibfnamefont {R.~M.}\ \bibnamefont {Young}}, \ and\ \bibinfo
  {author} {\bibfnamefont {M.~R.}\ \bibnamefont {Wasielewski}},\ }\href
  {\doibase 10.1021/acs.jpcc.1c09432} {\bibfield  {journal} {\bibinfo
  {journal} {The Journal of Physical Chemistry C}\ }\textbf {\bibinfo {volume}
  {126}},\ \bibinfo {pages} {120} (\bibinfo {year} {2022})}\BibitemShut
  {NoStop}%
\bibitem [{\citenamefont {Yan}\ and\ \citenamefont {Mukamel}(1986)}]{Yan1986a}%
  \BibitemOpen
  \bibfield  {author} {\bibinfo {author} {\bibfnamefont {Y.~J.}\ \bibnamefont
  {Yan}}\ and\ \bibinfo {author} {\bibfnamefont {S.}~\bibnamefont {Mukamel}},\
  }\href {\doibase 10.1063/1.451502} {\bibfield  {journal} {\bibinfo  {journal}
  {The Journal of Chemical Physics}\ }\textbf {\bibinfo {volume} {85}},\
  \bibinfo {pages} {5908} (\bibinfo {year} {1986})}\BibitemShut {NoStop}%
\bibitem [{\citenamefont {Fidler}\ and\ \citenamefont
  {Engel}(2013)}]{Fidler13a}%
  \BibitemOpen
  \bibfield  {author} {\bibinfo {author} {\bibfnamefont {A.~F.}\ \bibnamefont
  {Fidler}}\ and\ \bibinfo {author} {\bibfnamefont {G.~S.}\ \bibnamefont
  {Engel}},\ }\href {\doibase 10.1021/jp311713x} {\bibfield  {journal}
  {\bibinfo  {journal} {The Journal of Physical Chemistry A}\ }\textbf
  {\bibinfo {volume} {117}},\ \bibinfo {pages} {9444} (\bibinfo {year}
  {2013})}\BibitemShut {NoStop}%
\bibitem [{\citenamefont {Spano}(2010)}]{Spano2010a}%
  \BibitemOpen
  \bibfield  {author} {\bibinfo {author} {\bibfnamefont {F.~C.}\ \bibnamefont
  {Spano}},\ }\href {\doibase 10.1021/ar900233v} {\bibfield  {journal}
  {\bibinfo  {journal} {Accounts of Chemical Research}\ }\textbf {\bibinfo
  {volume} {43}},\ \bibinfo {pages} {429} (\bibinfo {year} {2010})}\BibitemShut
  {NoStop}%
\bibitem [{\citenamefont {Hayes}\ \emph {et~al.}(2013)\citenamefont {Hayes},
  \citenamefont {Griffin},\ and\ \citenamefont {Engel}}]{Hayes13a}%
  \BibitemOpen
  \bibfield  {author} {\bibinfo {author} {\bibfnamefont {D.}~\bibnamefont
  {Hayes}}, \bibinfo {author} {\bibfnamefont {G.~B.}\ \bibnamefont {Griffin}},
  \ and\ \bibinfo {author} {\bibfnamefont {G.~S.}\ \bibnamefont {Engel}},\
  }\href {\doibase 10.1126/science.1233828} {\bibfield  {journal} {\bibinfo
  {journal} {Science}\ }\textbf {\bibinfo {volume} {340}},\ \bibinfo {pages}
  {1431} (\bibinfo {year} {2013})}\BibitemShut {NoStop}%
\bibitem [{\citenamefont {Halpin}\ \emph {et~al.}(2014)\citenamefont {Halpin},
  \citenamefont {Johnson}, \citenamefont {Tempelaar}, \citenamefont {Murphy},
  \citenamefont {Knoester}, \citenamefont {Jansen},\ and\ \citenamefont
  {Miller}}]{Halpin2014}%
  \BibitemOpen
  \bibfield  {author} {\bibinfo {author} {\bibfnamefont {A.}~\bibnamefont
  {Halpin}}, \bibinfo {author} {\bibfnamefont {P.~J.}\ \bibnamefont {Johnson}},
  \bibinfo {author} {\bibfnamefont {R.}~\bibnamefont {Tempelaar}}, \bibinfo
  {author} {\bibfnamefont {R.~S.}\ \bibnamefont {Murphy}}, \bibinfo {author}
  {\bibfnamefont {J.}~\bibnamefont {Knoester}}, \bibinfo {author}
  {\bibfnamefont {T.~L.}\ \bibnamefont {Jansen}}, \ and\ \bibinfo {author}
  {\bibfnamefont {R.~J.}\ \bibnamefont {Miller}},\ }\href@noop {} {\bibfield
  {journal} {\bibinfo  {journal} {Nat. Chem.}\ }\textbf {\bibinfo {volume}
  {6}},\ \bibinfo {pages} {196} (\bibinfo {year} {2014})}\BibitemShut {NoStop}%
\bibitem [{\citenamefont {Ishizaki}\ \emph {et~al.}(2010)\citenamefont
  {Ishizaki}, \citenamefont {Calhoun}, \citenamefont {Schlau-Cohen},\ and\
  \citenamefont {Fleming}}]{Ishizaki10a}%
  \BibitemOpen
  \bibfield  {author} {\bibinfo {author} {\bibfnamefont {A.}~\bibnamefont
  {Ishizaki}}, \bibinfo {author} {\bibfnamefont {T.~R.}\ \bibnamefont
  {Calhoun}}, \bibinfo {author} {\bibfnamefont {G.~S.}\ \bibnamefont
  {Schlau-Cohen}}, \ and\ \bibinfo {author} {\bibfnamefont {G.~R.}\
  \bibnamefont {Fleming}},\ }\href {\doibase 10.1039/C003389H} {\bibfield
  {journal} {\bibinfo  {journal} {Phys. Chem. Chem. Phys.}\ }\textbf {\bibinfo
  {volume} {12}},\ \bibinfo {pages} {7319} (\bibinfo {year}
  {2010})}\BibitemShut {NoStop}%
\bibitem [{\citenamefont {Tiwari}\ \emph {et~al.}(2013)\citenamefont {Tiwari},
  \citenamefont {Peters},\ and\ \citenamefont {Jonas}}]{Tiwari13a}%
  \BibitemOpen
  \bibfield  {author} {\bibinfo {author} {\bibfnamefont {V.}~\bibnamefont
  {Tiwari}}, \bibinfo {author} {\bibfnamefont {W.~K.}\ \bibnamefont {Peters}},
  \ and\ \bibinfo {author} {\bibfnamefont {D.~M.}\ \bibnamefont {Jonas}},\
  }\href {\doibase 10.1073/pnas.1211157110} {\bibfield  {journal} {\bibinfo
  {journal} {Proceedings of the National Academy of Sciences}\ }\textbf
  {\bibinfo {volume} {110}},\ \bibinfo {pages} {1203} (\bibinfo {year}
  {2013})}\BibitemShut {NoStop}%
\bibitem [{\citenamefont {Butkus}\ \emph {et~al.}(2014)\citenamefont {Butkus},
  \citenamefont {Valkunas},\ and\ \citenamefont {Abramavicius}}]{Butkus14a}%
  \BibitemOpen
  \bibfield  {author} {\bibinfo {author} {\bibfnamefont {V.}~\bibnamefont
  {Butkus}}, \bibinfo {author} {\bibfnamefont {L.}~\bibnamefont {Valkunas}}, \
  and\ \bibinfo {author} {\bibfnamefont {D.}~\bibnamefont {Abramavicius}},\
  }\href {\doibase 10.1063/1.4861466} {\bibfield  {journal} {\bibinfo
  {journal} {The Journal of Chemical Physics}\ }\textbf {\bibinfo {volume}
  {140}},\ \bibinfo {pages} {034306} (\bibinfo {year} {2014})}\BibitemShut
  {NoStop}%
\bibitem [{\citenamefont {Krčmář}\ \emph {et~al.}(2015)\citenamefont
  {Krčmář}, \citenamefont {Gelin},\ and\ \citenamefont {Domcke}}]{krcmar}%
  \BibitemOpen
  \bibfield  {author} {\bibinfo {author} {\bibfnamefont {J.}~\bibnamefont
  {Krčmář}}, \bibinfo {author} {\bibfnamefont {M.~F.}\ \bibnamefont
  {Gelin}}, \ and\ \bibinfo {author} {\bibfnamefont {W.}~\bibnamefont
  {Domcke}},\ }\href {\doibase 10.1063/1.4928685} {\bibfield  {journal}
  {\bibinfo  {journal} {The Journal of Chemical Physics}\ }\textbf {\bibinfo
  {volume} {143}},\ \bibinfo {pages} {074308} (\bibinfo {year}
  {2015})}\BibitemShut {NoStop}%
\bibitem [{\citenamefont {Duan}\ and\ \citenamefont
  {Thorwart}(2016)}]{Duan2016}%
  \BibitemOpen
  \bibfield  {author} {\bibinfo {author} {\bibfnamefont {H.-G.}\ \bibnamefont
  {Duan}}\ and\ \bibinfo {author} {\bibfnamefont {M.}~\bibnamefont
  {Thorwart}},\ }\href {\doibase 10.1021/acs.jpclett.5b02793} {\bibfield
  {journal} {\bibinfo  {journal} {The Journal of Physical Chemistry Letters}\
  }\textbf {\bibinfo {volume} {7}},\ \bibinfo {pages} {382} (\bibinfo {year}
  {2016})}\BibitemShut {NoStop}%
\bibitem [{\citenamefont {Li}\ \emph {et~al.}(2021)\citenamefont {Li},
  \citenamefont {Ko}, \citenamefont {Yang}, \citenamefont {Sarovar},\ and\
  \citenamefont {Whaley}}]{Li2021a}%
  \BibitemOpen
  \bibfield  {author} {\bibinfo {author} {\bibfnamefont {Z.-Z.}\ \bibnamefont
  {Li}}, \bibinfo {author} {\bibfnamefont {L.}~\bibnamefont {Ko}}, \bibinfo
  {author} {\bibfnamefont {Z.}~\bibnamefont {Yang}}, \bibinfo {author}
  {\bibfnamefont {M.}~\bibnamefont {Sarovar}}, \ and\ \bibinfo {author}
  {\bibfnamefont {K.~B.}\ \bibnamefont {Whaley}},\ }\href {\doibase
  10.1088/1367-2630/abedfe} {\bibfield  {journal} {\bibinfo  {journal} {New
  Journal of Physics}\ }\textbf {\bibinfo {volume} {23}},\ \bibinfo {pages}
  {073012} (\bibinfo {year} {2021})}\BibitemShut {NoStop}%
\bibitem [{\citenamefont {Caycedo-Soler}\ \emph {et~al.}(2022)\citenamefont
  {Caycedo-Soler}, \citenamefont {Mattioni}, \citenamefont {Lim}, \citenamefont
  {Renger}, \citenamefont {Huelga},\ and\ \citenamefont
  {Plenio}}]{Caycedo-Soler2022}%
  \BibitemOpen
  \bibfield  {author} {\bibinfo {author} {\bibfnamefont {F.}~\bibnamefont
  {Caycedo-Soler}}, \bibinfo {author} {\bibfnamefont {A.}~\bibnamefont
  {Mattioni}}, \bibinfo {author} {\bibfnamefont {J.}~\bibnamefont {Lim}},
  \bibinfo {author} {\bibfnamefont {T.}~\bibnamefont {Renger}}, \bibinfo
  {author} {\bibfnamefont {S.~F.}\ \bibnamefont {Huelga}}, \ and\ \bibinfo
  {author} {\bibfnamefont {M.~B.}\ \bibnamefont {Plenio}},\ }\href {\doibase
  10.1038/s41467-022-30565-4} {\bibfield  {journal} {\bibinfo  {journal}
  {Nature Communications}\ }\textbf {\bibinfo {volume} {13}},\ \bibinfo {pages}
  {2912} (\bibinfo {year} {2022})}\BibitemShut {NoStop}%
\bibitem [{\citenamefont {Azumi}\ and\ \citenamefont
  {Matsuzaki}(1977)}]{Azumi}%
  \BibitemOpen
  \bibfield  {author} {\bibinfo {author} {\bibfnamefont {T.}~\bibnamefont
  {Azumi}}\ and\ \bibinfo {author} {\bibfnamefont {K.}~\bibnamefont
  {Matsuzaki}},\ }\href {\doibase
  https://doi.org/10.1111/j.1751-1097.1977.tb06918.x} {\bibfield  {journal}
  {\bibinfo  {journal} {Photochemistry and Photobiology}\ }\textbf {\bibinfo
  {volume} {25}},\ \bibinfo {pages} {315} (\bibinfo {year} {1977})}\BibitemShut
  {NoStop}%
\bibitem [{\citenamefont {Witkowski}\ and\ \citenamefont
  {Moffitt}(1960)}]{Witkowski}%
  \BibitemOpen
  \bibfield  {author} {\bibinfo {author} {\bibfnamefont {A.}~\bibnamefont
  {Witkowski}}\ and\ \bibinfo {author} {\bibfnamefont {W.}~\bibnamefont
  {Moffitt}},\ }\href {\doibase 10.1063/1.1731278} {\bibfield  {journal}
  {\bibinfo  {journal} {The Journal of Chemical Physics}\ }\textbf {\bibinfo
  {volume} {33}},\ \bibinfo {pages} {872} (\bibinfo {year} {1960})}\BibitemShut
  {NoStop}%
\bibitem [{\citenamefont {Mahan}(2000)}]{Mahan}%
  \BibitemOpen
  \bibfield  {author} {\bibinfo {author} {\bibfnamefont {G.~D.}\ \bibnamefont
  {Mahan}},\ }\href {https://doi.org/10.1017/CBO9780511675935} {\emph {\bibinfo
  {title} {Many-particle physics}}}\ (\bibinfo  {publisher} {Kuwer Academic,
  Boston},\ \bibinfo {year} {2000})\BibitemShut {NoStop}%
\bibitem [{\citenamefont {Breuer}\ \emph {et~al.}(2002)\citenamefont {Breuer},
  \citenamefont {Petruccione},\ and\ \citenamefont {Petruccione}}]{Breuer}%
  \BibitemOpen
  \bibfield  {author} {\bibinfo {author} {\bibfnamefont {H.}~\bibnamefont
  {Breuer}}, \bibinfo {author} {\bibfnamefont {F.}~\bibnamefont {Petruccione}},
  \ and\ \bibinfo {author} {\bibfnamefont {S.}~\bibnamefont {Petruccione}},\
  }\href {https://books.google.it/books?id=0Yx5VzaMYm8C} {\emph {\bibinfo
  {title} {The Theory of Open Quantum Systems}}}\ (\bibinfo  {publisher}
  {Oxford University Press},\ \bibinfo {year} {2002})\BibitemShut {NoStop}%
\end{thebibliography}
\end{document}